# Cooperation with an Untrusted Relay: A Secrecy Perspective

Xiang He Aylin Yener

*xxh119@psu.edu* *yener@ee.psu.edu*

Wireless Communications and Networking Laboratory

The Pennsylvania State University, University Park, PA 16802

**Abstract**

We consider the communication scenario where a source-destination pair wishes to keep the information secret from a relay node despite wanting to enlist its help. For this scenario, an interesting question is whether the relay node should be deployed at all. That is, whether cooperation with an untrusted relay node can ever be beneficial. We first provide an achievable secrecy rate for the general untrusted relay channel, and proceed to investigate this question for two types of relay networks with orthogonal components. For the first model, there is an orthogonal link from the source to the relay. For the second model, there is an orthogonal link from the relay to the destination. For the first model, we find the equivocation capacity region and show that answer is negative. In contrast, for the second model, we find that the answer is positive. Specifically, we show by means of the achievable secrecy rate based on compress-and-forward, that, by asking the untrusted relay node to relay information, we can achieve a higher secrecy rate than just treating the relay as an eavesdropper. For a special class of the second model, where the relay is not interfering itself, we derive an upper bound for the secrecy rate using an argument whose net effect is to separate the eavesdropper from the relay. The merit of the new upper bound is demonstrated on two channels that belong to this special class. The Gaussian case of the second model mentioned above benefits from this approach in that the new upper bound improves the previously known bounds. For the Cover-Kim deterministic relay channel, the new upper bound finds the secrecy capacity when the source-destination link is not worse than the source-relay link, by matching with achievable rate we present.

This work was presented in part at the 41st Asilomar Conference on Signals, System and Computers, November 2007, Information Theory and Application Workshop, ITA 2008, January, 2008, and International Symposium on Information Theory, ISIT 2008, July, 2008. This work is supported in part by the National Science Foundation with Grants CCR-0237727, CCF-051483, CNS-0716325, and the DARPA ITMANET Program with Grant W911NF-07-1-0028.

## I. Introduction

A fundamental approach to information security is founded in information theory where the limits of reliable communication can be determined while keeping the information secret from eavesdropping node(s). This notion of secrecy was first proposed by Shannon [1]. In his work, Shannon assumed that the eavesdropper has perfect access to the signal transmitted from the source to the destination and determined that the rate of key must equal to the rate of data to ensure “perfect secrecy”, i.e., in order for the data not to be leaked to the eavesdropper even if the eavesdropper has unlimited computational power. Wyner, in [2], pointed out that Shannon’s assumption is pessimistic, as more often than not, the eavesdropper only has a noisy copy of the signal transmitted from the source and building a useful secure communication system per Shannon’s notion is possible [2], [3].

Recent work in this area aims to find the secrecy capacity or capacity region for a variety of communication scenarios and channel models. A set of models follows the classical model of Wyner’s wiretap channel [2], where an external eavesdropper is present in addition to legitimate parties. This line of work includes the multiple input multiple output (MIMO) wiretap channel [4]–[6], the wiretap channel with a cooperative jammer [7], the multiple access wiretap channel [8]–[11], the MIMO broadcast channel with multiple legitimate receivers and an external eavesdropper [12], the two-way wiretap channel [10], the relay channel with an external eavesdropper [13]. In these models, where the information leaked to the eavesdropper is a loss to the legitimate communication system, it was observed that legitimate parties could aid in enhancing secrecy by introducing intentional interference to the eavesdropper via cooperative jamming [7], [10], [14]. Another set of models deals with a more symmetric scenario, where each receiver of an intended message is also modeled as an eavesdropper for the remaining unintended messages in the system. This setting has been considered for the multiple access [15], broadcast [16], [17], and interference channels [18], [19]. In these models, one communication pair, in the interest of protecting its own information, may end up helping the other pair [19].

The focus of this work, while on cooperative communications, differs from the above models in that, it deals with a communication network whose nodes have different levels of security clearance. Examples like this exist in real life. In a government intelligence network or the network of a financial institution, not every node in the network is supposed to have the same level of access to information, despite operating with agreed protocols and serving as relay nodes in the network. The question is whether these *untrusted* nodes should still participate in this cooperative communication network, or if they pose a "problem" when secret messages are to be transmitted, and hence, their cooperation should not be enlisted. The basic issue addressed here therefore, different from the previous models that aim to solve the co-existence problem of several communication pairs, is to resolve the conflict within one system.

This paper focuses on the most basic model in this category in order to assess the effect of secrecy requirements upon cooperative communications. We consider the three node relay network, where the relay has a lower security clearance than the destination and is therefore untrusted. Reference [20], the first work that studied this model, shows that the secrecy capacity of this system is zero if the relay channel is degraded. The secrecy capacity equals that of the wiretap channel if the channel is reversely degraded, which means that the relay-to-destination communication is useless in this case as well [21]. In Section IV of this paper, we present yet another negative result: the relay node is again useless in a class of relay networks with orthogonal components [22], dubbed Model $1$ in the sequel. In each of these references, the model turns out to be equivalent to the models of the first two categories, where the relay node is merely an eavesdropper, rather than a cooperating partner.

In light of these results, one might be tempted to take a pessimistic view, and wonder whether there exists any situation where the cooperation of the untrusted relay might enable a higher secrecy rate than simply treating it as an eavesdropper. Interestingly, we find in this paper that the answer to this question is yes. This is shown for a class of relay networks with orthogonal components where the relay to destination link is orthogonal to that from the source [23], dubbed Model $2$ in the sequel. Specifically, we observe that by performing compress-and-forward, the relay node can help increase an otherwise zero secrecy rate without having any idea what it is relaying.

Once an achievable secrecy rate with the untrusted relay being useful is found, an upper bound on the secrecy rate is needed to assess how close the achievable strategy is to the optimum. There are two previously known upper bounds. Reference [13] provided an upper bound for the relay channel with an external eavesdropper. By assuming that this external eavesdropper receives the signals received and transmitted by the relay, we observe the model in [13] can be specialized into the model considered in this work and hence the bound in [13] can be readily applied. Alas, this bound is not computable for the Gaussian case. A computable bound was provided for the Gaussian relay channel with a co-located eavesdropper in [21]. Alas, this bound does not depend on the condition of the relay-to-destination channel. Moreover, the noise correlation of the links may render the bound to be arbitrarily loose. In this work, we aim to derive an upper bound that improves the bound in [21] in these two aspects, and accomplish this goal for a class of untrusted relay channels.

More specifically, the upper bound on the secrecy rate is derived for a special class of Model 2, where the relay is not interfering itself. The derivation of the upper bound entails the introduction of a second eavesdropper. Although in general, introducing a second eavesdropper can decrease the secrecy capacity, we prove that for the special class of channels question, doing so does not alter the secrecy capacity. The upper bound is then derived by removing the first eavesdropper at the relay and introducing correlation between the output seen by the second eavesdropper and other outputs of the channel, which tightens the upper bound as in other Sato-type bounds; see [5] for example.

The merit of the new upper bound is demonstrated in two cases: First, for the Gaussian case of Model 2, we show that the new bound improves the previously known bounds. Second, for the Gaussian Cover-Kim deterministic relay channel introduced in [24], we show that the upper bound matches the achievable rate using compress-and-forward when the signal to noise ratio of the source-destination link is not worse than that of the source-relay link, thus, establishing the secrecy capacity.

The remainder of the paper is organized as follows: Section II describes the general relay network with a co-located eavesdropper, and an achievable equivocation region for this channel using the compress-and-forward relaying. In section III, the two special cases of the general model, i.e., Model 1 and Model 2 are described. Section IV presents the equivocation capacity region for Model 1. Section V specializes the achievable region found in Section II to Model 2. Section VI identifies a special class of Model 2, for which introducing a second eavesdropper properly will not decrease the secrecy capacity, and derives an upper bound for its secrecy rate. The upper bound is then specialized to the Gaussian case of Model 2. Section VII investigates the secrecy capacity of Gaussian Cover-Kim deterministic relay channel. We note that to facilitate a better flow throughout the manuscript, more involved proofs are presented in appendices whereas shorter ones are kept in the main text.

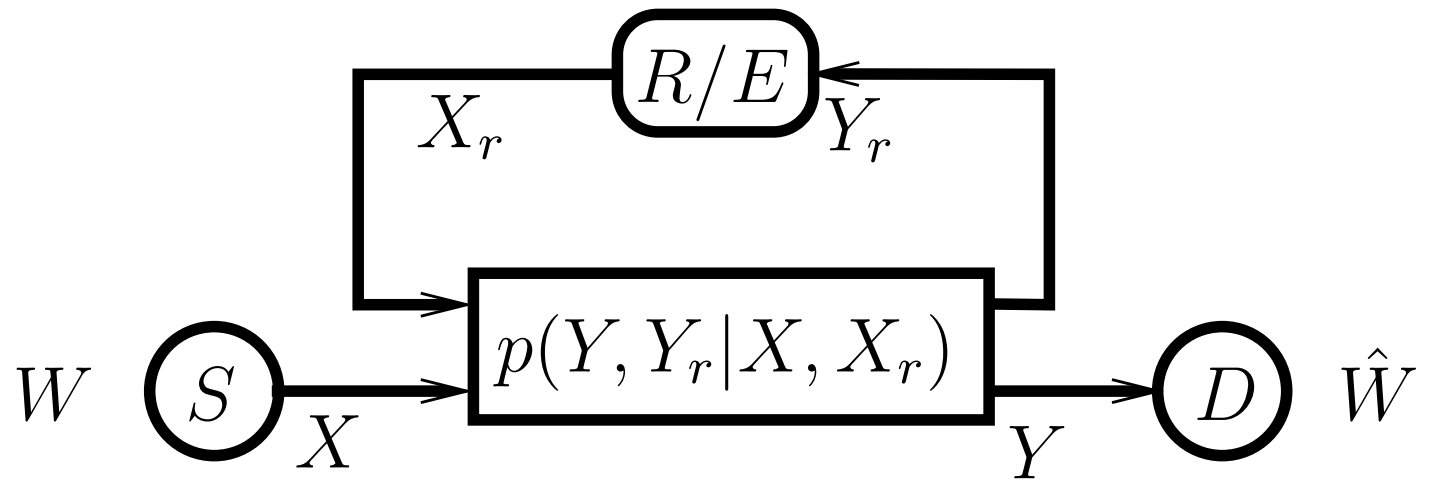

Fig. 1. Relay Channel with a Co-located Eavesdropper: S: Source Node, R/E: Relay Node with a Co-located Eavesdropper, D: Destination Node.

Throughout this paper, the following notation is used: $\forall k$, $\varepsilon_k$ denotes a variable that goes to 0 when $n$ goes to $\infty$. $C(x) = \frac{1}{2}\log_2(1+x)$. $X^n$ denotes a vector of length $n$, whereas $X_i$ denotes the $i$th element of the vector. $X_{]i[}$ denotes the set $\{X_j, 1 \leq j < i \text{ or } i < j \leq n\}$. $X_1^i$ denotes the set $\{X_j, 1 \leq j \leq i\}$; the set is empty if $i < 1$. $\lfloor a \rfloor$ denotes the largest integer less than or equal to $a$. The short hand $W(a, ..., b)$ stands for the set $\{W(a), W(a+1), ..., W(b)\}$. The short hand $W^a$ stands for the set $\{W(1), W(2), ..., W(a)\}$.

## II. Achievable Secrecy Rate for the General Relay Channel with Co-located Eavesdropper

The relay channel with a co-located eavesdropper was first considered in [20] and is shown in Figure 1. It is a memoryless three-node relay channel [25], whose description is $p(Y, Y_r|X, X_r)$. $X, X_r$ are the channel inputs from the source and the relay respectively, and $Y, Y_r$ are the channel outputs observed by the destination and the relay respectively. We assume that there is an eavesdropper at the relay node who has access to everything that the relay node knows. The source wishes to send message $W$ to the destination over $n$ channel uses, while keeping it secret from the eavesdropper.

Without loss of generality, the relaying function for the $i$th channel use can be defined as

$$X_{r,i} = g_i\left(X_r^{i-1}, Y_r^{i-1}, A\right) \tag{1}$$

where $A$ is a random variable which models any stochastic mapping employed by the relay node. Hence, without loss of generality, we can restrict $g_i$ to be a deterministic function.

The information available to the eavesdropper regarding the secret message $W$ is $\{X_r^n, Y_r^n, A\}$.

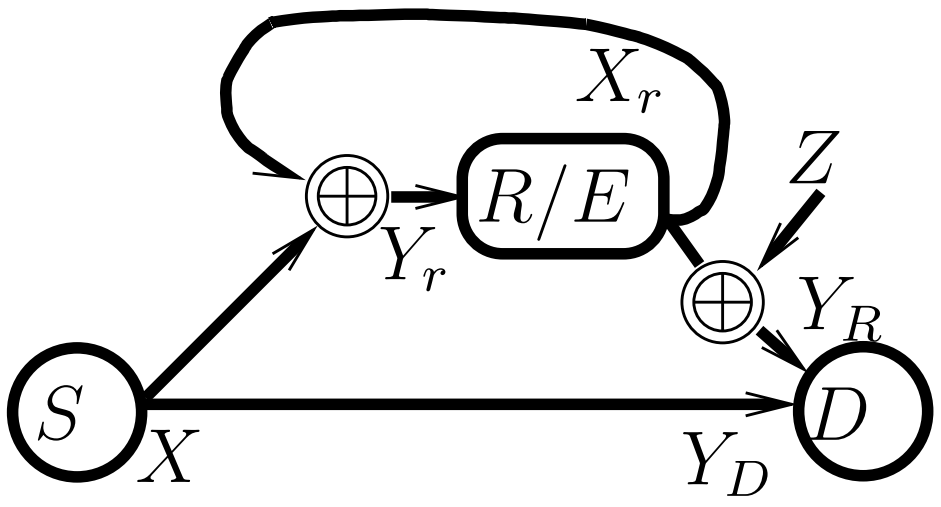

Fig. 2. Example: Relay's transmitted signal $X_r^n$ can provide more information about $W$

Thus, the equivocation rate is computed as

$$R_e = \lim_{n\to\infty} \frac{1}{n} H\left(W|X_r^n, Y_r^n, A\right) \tag{2}$$

Note that since $W - \{X_r^n, Y_r^n\} - A$ is a Markov chain, we have $H\left(W|X_r^n Y_r^n A\right) = H\left(W|X_r^n Y_r^n\right)$. Hence the equivocation rate can be instead defined as

$$R_e = \lim_{n\to\infty} \frac{1}{n} H\left(W|X_r^n, Y_r^n\right) \tag{3}$$

.

*Remark 1:* In general $H(W|Y_r^n) > H(W|X_r^n, Y_r^n)$. To see an example, consider the channel model in Figure 2. Let $\oplus$ denote the binary addition. The channel has binary input. The destination receives signals from two orthogonal links: $Y_R$ and $Y_D$, and that we have

$$Y_D = X \tag{4}$$

$$Y_R = X_r \oplus Z \tag{5}$$

where $Z$ represents the i.i.d. binary noise.

It follows in this setting that, $H(W|X_r^n, Y_r^n) = 0$. This is because the relay can always subtract the interference caused by $X_r^n$ on its received signal and hence obtains $X^n$. However, $H(W|Y_r^n) = H(W)$ if $X_r^n$ is chosen to be an i.i.d. binary sequences, each component of which takes the value $1$ with probability $\frac{1}{2}$. Thus, in this case, we clearly have $H(W|Y_r^n) > H(W|X_r^n Y_r^n)$. $\square$

We should note however that, $H(W|Y_r^n) = H(W|X_r^n, Y_r^n)$ if the relaying scheme is deterministic: $X_{r,i} = g_i(X_r^{i-1}, Y_r^{i-1})$. Also, note that, clearly, any outer bound derived for the equivocation $H(W|Y_r^n)$ is an outer bound for $H(W|X_r^n, Y_r^n)$.

With this preparation, the equivocation rate region can be defined as follows: Let the message decoded by the destination be $\hat{W}$. The equivocation rate region is composed of all rate pairs $(R_1, R_e)$ such that:

$$
\begin{aligned}
&R_1 = \lim_{n\to\infty} \frac{1}{n} \log_2 |W| \\
&R_e = \lim_{n\to\infty} \frac{1}{n} H\left(W|X_r^n, Y_r^n\right) \\
&\text{s.t. } \lim_{n\to\infty} \frac{1}{n} \Pr\left(W \neq \hat{W}\right) = 0
\end{aligned}
$$

Here $|W|$ is the cardinality of the message set $W$. Note that when block Markov coding scheme [25] is used, the message is transmitted via successive blocks. In this case, $W$ denotes the messages transmitted over all blocks. $n$ should be the total number of channel uses of these blocks. The definition of $X_r^n, Y_r^n$ should be adjusted accordingly.

Next, we derive an achievable equivocation region based on compress-and-forward. Compress-and-forward scheme was proposed in [25] and has been used for the relay network with an external eavesdropper in [13], [26]. In our case, as we will see, the fact that the relay and the eavesdropper being co-located brings additional advantage to allow for a higher degree of compression to be achieved at the relay as compared to the setting in [13].

*Theorem 1:* For a relay network described as $p(Y, Y_r|X, X_r)$, with $X$, $X_r$ being the input from the source and the relay respectively, and $Y_r$,$Y$ being the signals received by the relay and the destination respectively, the following region of rate pairs $(R_1, R_e)$ is achievable.

$$
\bigcup \left\{ \begin{array}{rl} R_e \leq R_1 < & I\left(X; Y\hat{Y}_r|X_r\right) \\ 0 \leq R_e < & \left[I\left(X; Y\hat{Y}_r|X_r\right) - I\left(X; Y_r|X_r\right)\right]^+ \end{array} \right\} \tag{6}
$$

where

$$
I(X_r; Y) > I(\hat{Y}_r; Y_r|Y X_r) \tag{7}
$$

and the union is taken over:

$$
p(X)p(X_r)p(Y, Y_r|X, X_r)p(\hat{Y}_r|Y_r, X_r) \tag{8}
$$

*Proof:* See Appendix A. ■

*Remark 2:* Compared with the coding scheme presented in [13], the difference is that we have Wyner-Ziv coding. Without Wyner-Ziv coding, the constraint (7) in the Theorem would be

$$I(X_r;Y) > I(\hat{Y}_r;Y_r|X_r) \tag{9}$$

which is identical to that in [13, Theorem 4 (12)]. In [13], the eavesdropper is external to the relay node, and hence only has a noisy copy of $X_r^n(k)$. In this case, the equivocation over multiple blocks would not necessarily be the sum of equivocation over each block. Reference [13] worked around this problem by using a compress-and-forward scheme without Wyner-Ziv coding. The equivocation over multiple blocks was then lower bounded by proving that given the signal received by the eavesdropper and the secret message, the external eavesdropper would be able to determine the signals transmitted by the source and the relay via backward decoding [13, Appendix D (53)-(55)].

In contrast to that in [13], fortunately, in our model, the eavesdropper has perfect knowledge of $X_r^n(k)$. This enables us to compute the equivocation of $N$ blocks from the equivocation of each block. See (137)-(139) in Appendix A. Hence, the Wyner-Ziv coding is used in our setting without difficulty. □

*Remark 3:* Theorem 1 will be useful in Section V in finding an achievable rate for one of the models (Model $2$) that we will describe in the next section. □

*Remark 4:* We can prefix the channel input $X$ with $U$ and apply Theorem 1 to the channel $p(Y,Y_r|U,X_r)$. The equivocation region then becomes:

$$\bigcup \left\{ \begin{array}{rl} R_e \leq R_1 < & I\left(U;Y\hat{Y}_r|X_r\right) \\ 0 \leq R_e < & \left[I\left(U;Y\hat{Y}_r|X_r\right) - I\left(U;Y_r|X_r\right)\right]^+ \end{array} \right\} \tag{10}$$

for which (7) must be fulfilled, and the union is taken over:

$$p(U,X)p(X_r)p(Y,Y_r|X,X_r)p(\hat{Y}_r|Y_r,X_r) \tag{11}$$

Clearly, this may potentially enlarge the achievable region given by Theorem 1. □

Having examined the general relay channel with a co-located eavesdropper, we next consider two special cases of it for which stronger results can be derived.

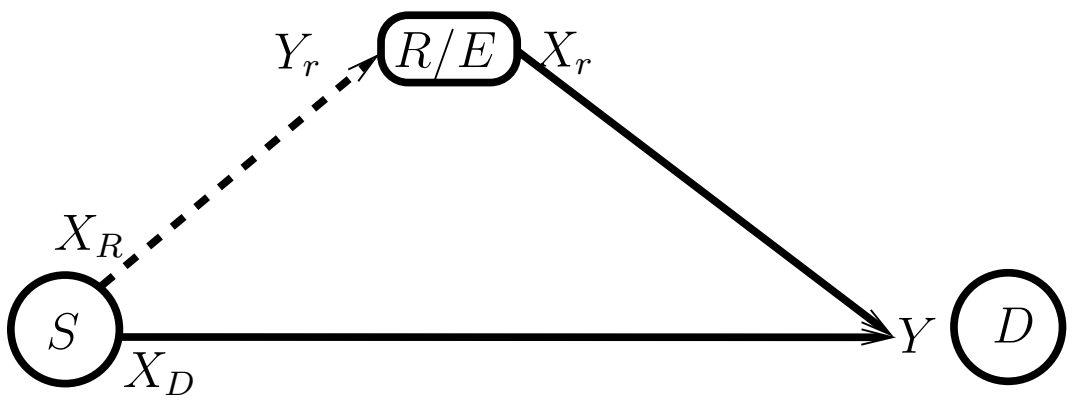

Fig. 3. Relay Channel with Orthogonal Components Model 1: Orthogonal Source to Relay Link

## III. Two Special Cases of the General Model: Relay Networks with Orthogonal Components

The two models of the relay network with orthogonal components are depicted in Figures 3 and 4 respectively. Figure 3 shows Model $1$. In this model, the relay and the source communicate with the destination via a multiple access channel, with its input being $X_D, X_r$ and output being $Y$. The source and the relay communicate via a channel orthogonal to the channel used by the source and the relay to transmit to the destination. The input and the output of this channel are denoted by $X_R$ and $Y_r$ respectively. Thus, the overall channel description is:

$$p\left(Y, Y_r | X_R, X_D, X_r\right) = p\left(Y | X_D, X_r\right) p\left(Y_r | X_R, X_r\right) \tag{12}$$

The capacity of this network without secrecy constraints was found in [22].

The Gaussian case of Model $1$ is defined as [22]:

$$Y_r = aX_R + Z_1, \quad Y = bX_r + X_D + Z \tag{13}$$

where $Z_1$ and $Z$ are independent zero mean real Gaussian random variables each with variance $N$. $a$ and $b$ are channel gains. The transmit power constraints on the source and the relay are given by:

$$\frac{1}{n}\sum_{i=1}^{n}\left(E[X_{R,i}^2] + E[X_{D,i}^2]\right) \leq P, \quad \frac{1}{n}\sum_{i=1}^{n} E[X_{r,i}^2] \leq \gamma P \tag{14}$$

Figure 4 shows Model $2$. In Model $2$, the source communicates with the relay and the destination via a broadcast channel, and the relay communicates with the destination via a separate (orthogonal) link. Thus, the channel is described by:

$$p\left(Y_D, Y_R, Y_r | X, X_r\right) = p(Y_D|X)p(Y_r|X, X_r, Y_D)p\left(Y_R | X_r\right) \tag{15}$$

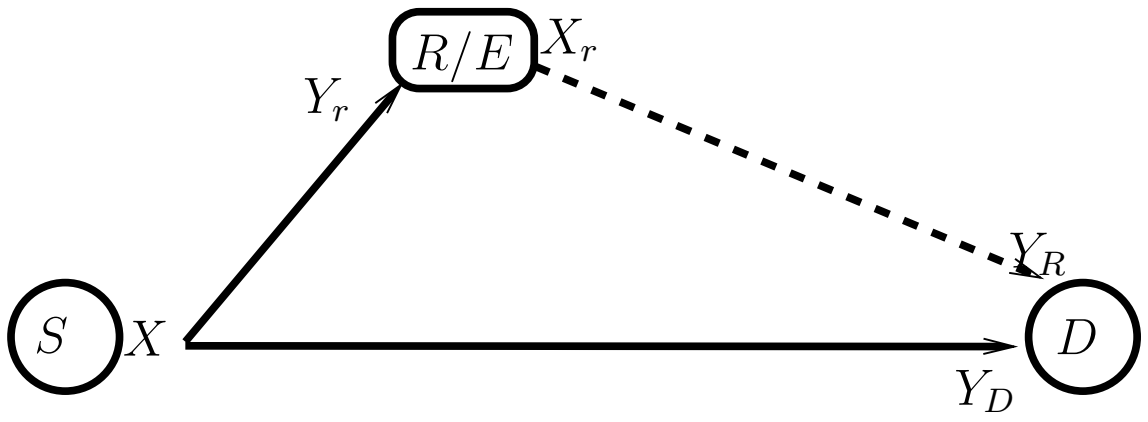

Fig. 4. Relay Channel with Orthogonal Components Model 2: Orthogonal Relay to Destination Link

When there are no secrecy constraints, the Gaussian case of Model $2$ was considered in [23]. The capacity of this channel remains an open problem except for some special cases given in [23].

The class of channels for which we will be able to derive an upper bound on the secrecy rate, is described by:

$$p(Y_D, Y_R, Y_r|X, X_r) = p(Y_D|X)p(Y_r|X, Y_D)p(Y_R|X_r) \tag{16}$$

Observe that such a channel is a special case of (15) since $X_r$ is dropped from the condition term of $Y_r$ in (15).

We will discuss two channels that fall into the class defined by (16): (i) the Gaussian case of Model $2$, (ii) the Gaussian Cover-Kim deterministic relay channel [24].

The Gaussian case of Model $2$ is defined as:

$$\begin{aligned} &Y_D = X + Z_D \quad Y_r = aX + Z_r \\ &Y_R = bX_r + Z_R \end{aligned} \tag{17}$$

where $Z_D, Z_r, Z_R$ are independent zero-mean Gaussian random variables with unit variance. $a$ and $b$ are channel gains. The transmit power of the source and the relay are constrained by:

$$\frac{1}{n}\sum_{i=1}^{n} E[X_{r,i}^2] \leq P_r, \quad \frac{1}{n}\sum_{i=1}^{n} E[X_i^2] \leq P \tag{18}$$

The Gaussian Cover-Kim deterministic relay channel is depicted in Figure 5. The received signals at the destination and at the relay are given by:

$$Y_D = X + Z, \quad Y_r = \alpha X - Z \tag{19}$$

where $\alpha$ is the channel gain and $Z$ is a zero mean Gaussian random variable with unit variance. Notice that the random variables representing the noise components have a correlation $\rho = -1$.

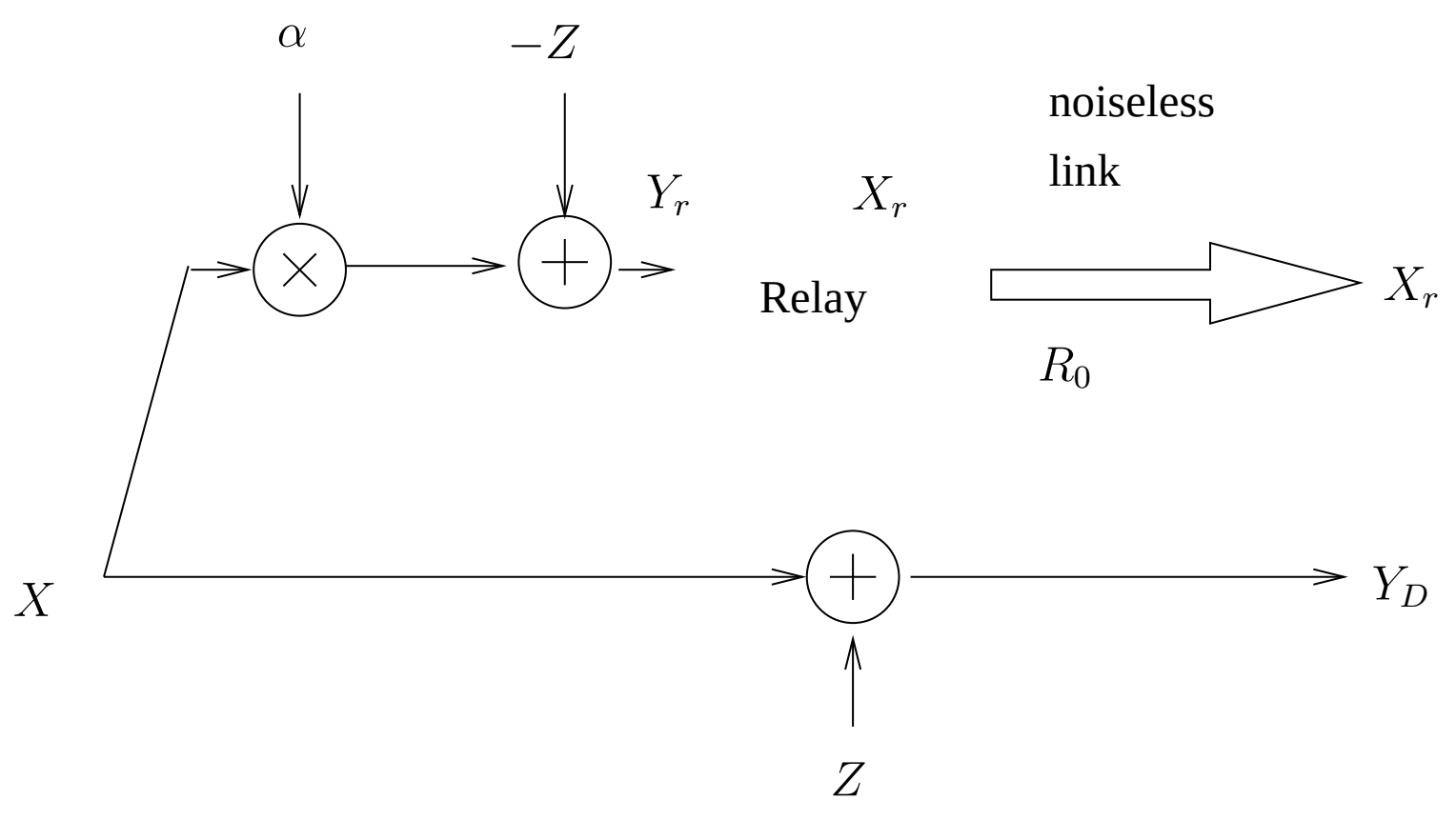

Fig. 5. The Gaussian Cover-Kim Deterministic Relay channel

Between the relay and the destination, there is a separate noiseless link with rate $R_0$. The destination receives side information from the relay via this link in addition to $Y_D$ which it receives from the source. The transmission power of the source is constrained to be:

$$\frac{1}{n}\sum_{i=1}^{n} E[X_i^2] \leq P \tag{20}$$

In the following sections, we first derive the equivocation capacity region of Model $1$. We then derive the achievable equivocation region for Model $2$ using the results from Section II. Finally, we derive the upper bound for the secrecy rate for the class of Model $2$ defined in (16) and specialize it to the Gaussian case and the Cover-Kim channel.

## IV. Equivocation Capacity Region for Model 1

*Theorem 2:* The equivocation capacity region of Model $1$ is given by

$$\bigcup_{\substack{p(X_r)\\ p(X_D|X_r)\\ p(X_R|X_r)}} \left\{ \begin{array}{l} (R_1, R_e): \\ 0 \leq R_1 < \min\left\{ I\left(X_D, X_r; Y\right), I\left(X_R; Y_r|X_r\right) + I\left(X_D; Y|X_r\right) \right\} \\ 0 \leq R_e < \min\{I\left(X_D; Y|X_r\right), R_1\} \end{array} \right\} \tag{21}$$

*Proof:* See Appendix B. ∎

*Remark 5:* Theorem 2 is proved by specializing the results from [21]. The achievable scheme is based on partial-decode-and-forward. This entails that the relay decodes the information transmitted via $X_R$. The scheme is outlined next for the sake of completeness:

Denote the codebook used by the relay and the source as $\mathcal{C}_r$ and $\{\mathcal{C}_R, \mathcal{C}_D\}$ respectively. The codeword in $\mathcal{C}_r$ is denoted by $X_r^n$. The codeword in $\{\mathcal{C}_R, \mathcal{C}_D\}$ is denoted by $\{X_D^n, X_R^n\}$, which are to be transmitted via $X_D$ and $X_R$ respectively.

The codebooks are generated as follows: $2^{n\min\{I(X_r;Y),I(X_R;Y_r|X_r)\}}$ codewords are sampled in an i.i.d. fashion from $p(X_r)$ to form $\mathcal{C}_r$. For each $X_r^n$ in $\mathcal{C}_r$, $2^{nI(X_D;Y|X_r)}$ codewords are sampled in an i.i.d. fashion from $p(X_D|X_r)$ and are included in $\mathcal{C}_D$. For each $X_r^n$ in $\mathcal{C}_r$, $2^{n\min\{I(X_r;Y),I(X_R;Y_r|X_r)\}}$ codewords are sampled in an i.i.d. fashion from $p(X_R|X_r)$ and are included in $\mathcal{C}_R$.

The transmission is divided into $N$ blocks, each composed of $n$ channel uses. The messages transmitted by the source during the $k$th block is denoted by $\{W_D(k), W_R(k)\}$. $W_D(k)$ corresponds to the secret part of the message. The cardinality of $\{W_D(k)\}$ is smaller than $2^{nI(X_D;Y|X_r)}$. The cardinality of $\{W_R(k)\}$ is smaller than $2^{n\min\{I(X_r;Y),I(X_R;Y_r|X_r)\}}$. The signals received and transmitted by the relay during the $k$th block are denoted by $Y_r^n(k)$ and $X_r^n(k)$ respectively. The relay decodes $W_R(k)$ from $Y_r^n(k)$ using $X_r^n(k)$ as the side information. $X_r^n(k)$ is chosen by the relay based on $W_R(k-1)$, which the relay decodes from $Y_r^n(k-1)$. The source node knows $W_R(k-1)$, and hence knows $X_r^n(k)$ before the $k$th block starts. It locates the part of the codebook $\mathcal{C}_R$ which is generated according to $X_r^n(k)$ and transmits the message $W_R(k)$ using this part of the codebook. The source also locates the part of the codebook $\mathcal{C}_D$ which is generated according to $X_r^n(k)$ and transmits the message $W_D(k)$ using this part of the codebook. The destination can successfully decode $W_R(k-1)$ from $Y^n(k)$, which determines $X_r^n(k)$, due to the fact that the cardinality of $\{W_R(k)\}$ is smaller than $2^{nI(X_r;Y)}$. Then it locates the part of the codebook in $\mathcal{C}_D$ that is generated according to $X_r^n(k)$ and use it to decode $W_D(k)$ from $Y^n(k)$. This is possible due to the fact that the cardinality of $\{W_D(k)\}$ is smaller than $2^{nI(X_D;Y|X_r)}$.$\square$

*Remark 6:* By letting $R_e = R_1$ in (21), we obtain the secrecy capacity of the network given by (22).

$$S = \max_{p(X_r)p(X_D|X_r)} I\left(Y; X_D|X_r\right) \tag{22}$$

$$= \max_{p(X_D|X_r=x_r)} I\left(Y; X_D|X_r = x_r\right) \tag{23}$$

It is readily seen that in this case the relay to destination link is not useful. Additionally, when $R_e < R_1$, from the coding scheme outlined in Remark 5, the secret information, $W_D(k)$, is only mapped to signal transmitted via $X_D$, which means the secret information does not pass through the relay node at all. These two observations combined lead to the conclusion that the relay-to-destination link is indeed *not useful* in improving the secrecy rate of the system, and that the untrusted relay should not be deployed at all. □

A direct extension of the above result can be readily made to the Gaussian channel. [1]

*Corollary 1:* For the Gaussian relay network described above, the equivocation region is given by (24).

$$\bigcup_{0\leq v,\rho\leq 1}\left\{\begin{array}{l} R_1 < \min\left\{ C\left(\frac{(v+b^2\gamma+2b\rho\sqrt{v\gamma})P}{N}\right), C\left(\frac{a^2(1-v)P}{N}\right)+C\left(\frac{v\left(1-\rho^2\right)P}{N}\right)\right\} \\ 0\leq R_e < C\left(\frac{v\left(1-\rho^2\right)P}{N}\right) \quad R_e\leq R_1 \end{array}\right\} \tag{24}$$

*Proof:* The proof is the same as in reference [22, Section III]. The three terms: $I(X_D, X_r; Y)$, $I(X_R; Y_r|X_r)$, $I(X_D; Y|X_r)$ are maximized simultaneously when $X_r, X_D, X_R$ are chosen to be zero mean and jointly Gaussian with the following parameters: $Var[X_r] = \gamma P, Var[X_R] = (1-v)P, Var[X_D] = vP, E[X_rX_D] = \rho P\sqrt{v\gamma}, E[X_RX_D] = 0$. ■

## V. An Achievable Region for Model 2

In this section, we present the achievable equivocation rate region for Model $2$.

*Theorem 3:* For Model $2$ defined by (15), an achievable equivocation rate region is given by:

$$\bigcup\left\{\begin{array}{rl} R_e\leq R_1 < & I\left(X; Y_D\hat{Y}_r|X_rY_R\right) \\ 0\leq R_e < & \left[I\left(X; Y_D\hat{Y}_r|X_rY_R\right) - I\left(X; Y_r|X_r\right)\right]^+ \end{array}\right\} \tag{25}$$

where

$$I(X_r; Y_R) > I(\hat{Y}_r; Y_r|Y_DY_RX_r) \tag{26}$$

and the union is taken over:

$$p(X)p(X_r)p(Y_D|X)p(Y_r|X, X_r, Y_D)p(Y_R|X_r)p(\hat{Y}_r|Y_r, X_r) \tag{27}$$

*Proof:* We use Theorem 1. In particular, region (25) follows from (6) by letting $Y = \{Y_D, Y_R\}$ and using the following two Markov chains (28) and (29). (29) follows from the fact that $\{X_r, Y_R\}$ is independent from $Y_D$ as shown by (27).

$$X - X_r - Y_R \tag{28}$$

[1] Proofs follow by replacing entropy with differential entropy whenever necessary.

$$X_r - Y_R - Y_D \tag{29}$$

It then follows from (28) and (29) that $I(X_r; Y_R Y_D) = I(X_r; Y_R)$ and

$$I(X; Y_R Y_D \hat{Y}_r | X_r) = I(X; Y_D \hat{Y}_r | X_r Y_R) \tag{30}$$

■

Next, we apply Theorem 3 to the Gaussian case, which is defined by (17).

*Corollary 2:* For the Gaussian relay network with orthogonal components defined by (17), the following rate region is achievable.

$$\bigcup_{0 \le p \le P} \left\{ \begin{array}{ll} (R_e, R_1): & 0 \le R_e \le R_1 < C\left(p + \frac{a^2 p}{1+\sigma_Q^2}\right) \\ & R_e < C\left(p + \frac{a^2 p}{1+\sigma_Q^2}\right) - C\left(a^2 p\right) \end{array} \right\} \tag{31}$$

where

$$\sigma_Q^2 = \frac{\left(a^2+1\right)p+1}{b^2 P_r \left(p+1\right)} \tag{32}$$

*Proof:* Region (31) follows from letting $X \sim \mathcal{N}(0,p), X_r \sim \mathcal{N}(0,P_r), \hat{Y}_r = Y_r + Z_Q, Z_Q \sim \mathcal{N}(0,\sigma_Q^2)$, and $Z_Q$ is independent from all the other variables. Substituting the distribution of $X, X_r, \hat{Y}_r, Y_r, Y_D, Y_R$ into (26), we find that we need

$$\sigma_Q^2 > \frac{\left(a^2+1\right)p+1}{b^2 P_r \left(p+1\right)} \tag{33}$$

It is clear from (31) that to make the region as large as possible, $\sigma_Q^2$ should be as small as possible, and (32) ensures this. ■

*Remark 7:* Suppose $a > 1$. Without the channel between relay and destination, we have a wiretap channel where the eavesdropper has a better channel. Hence, the secrecy capacity is zero [27]. We also know that a non-zero secrecy rate cannot be achieved with decode-and-forward. However, if the relay to destination gain, $b$, is large enough, a non-zero secrecy rate can be achieved with compress-and-forward, as can be seen from (31). This is an example where the relay-to-destination link helps to achieve a non-zero secrecy rate when the relay and the eavesdropper are co-located. Thus, the untrusted relay *is useful and should be cooperated with*. □

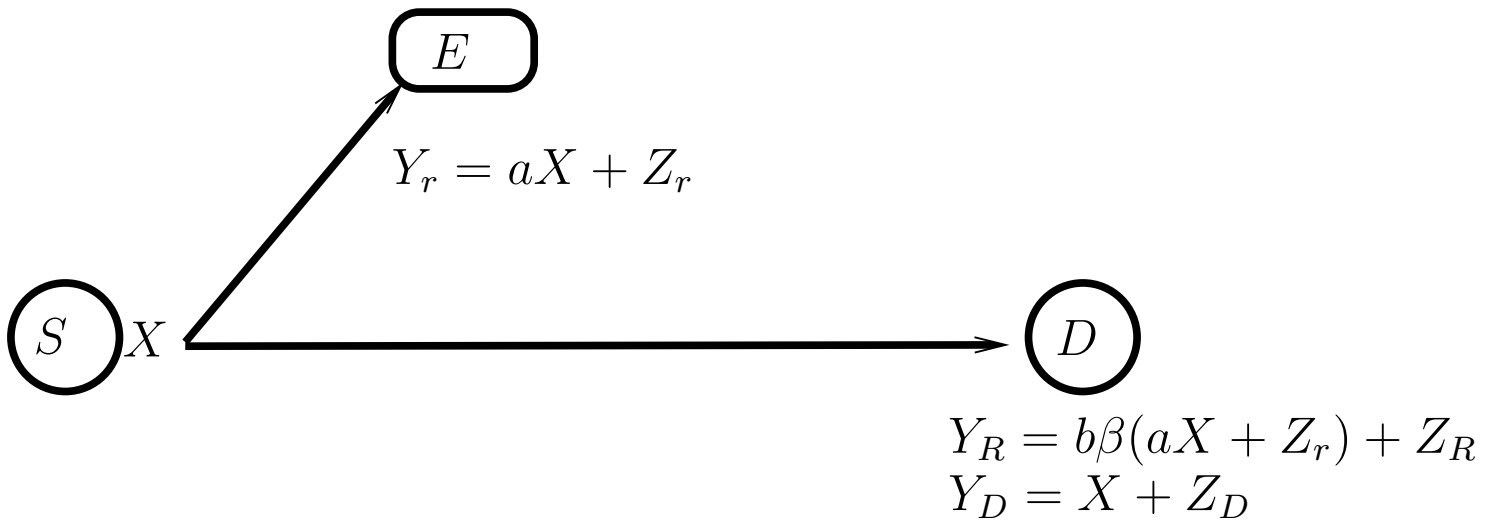

Fig. 6. The Equivalent Wiretap Channel of Model 2 using Amplify-and-forward Relaying

*Remark 8:* The scheme we present here differs from the noise forwarding scheme of [13] where the relay transmits noise that is independent from its received signal. By contrast, in this work, the signal transmitted by the relay is computed from its received signal. □

*Remark 9:* The amplify and forward scheme can also be used at the relay. Let $p$ be the average transmission power of the source node. Then, in this case, the signal transmitted by the relay at the $i$th channel use is given by

$$X_{r,i} = \beta Y_{r,i-1}, \qquad \text{where } \beta = \frac{\sqrt{P_r}}{\sqrt{a^2 p + 1}} \tag{34}$$

Note that in (34), we force $X_{r,i}$ to depend on the signal received in the previous channel use $Y_{r,i-1}$ in order to preserve the causality of the relay function as defined in (1). However, because the channel between the relay and the destination is orthogonal to the one between the source and the destination, the fact that the signals received via $Y_R$ is delayed by one channel use compared to those received via $Y_D$ does not make any difference to the destination. Therefore, it is safe to write $X_r = \beta Y_r$ and omit the subscript $i$.

The relay network is therefore equivalent to a Gaussian wiretap channel as shown in Figure 6. The achievable secrecy rate is computed from $[I(X;Y_R Y_D) - I(X;Y_r)]^+$ [3] for a Gaussian distribution for $X$: $X \sim \mathcal{N}(0,p)$ and when maximized over $p$, the secrecy rate is given by:

$$R_e < \max_{0\le p\le P} \frac{1}{2}\left[\log\left(1 + (1+\xi)\,p\right) - \log\left(1 + a^2 p\right)\right]^+ \tag{35}$$

where for $\beta$ defined in (34), $\xi$ is given by

$$\xi = \frac{a^2\beta^2 b^2}{1+\beta^2 b^2} \tag{36}$$

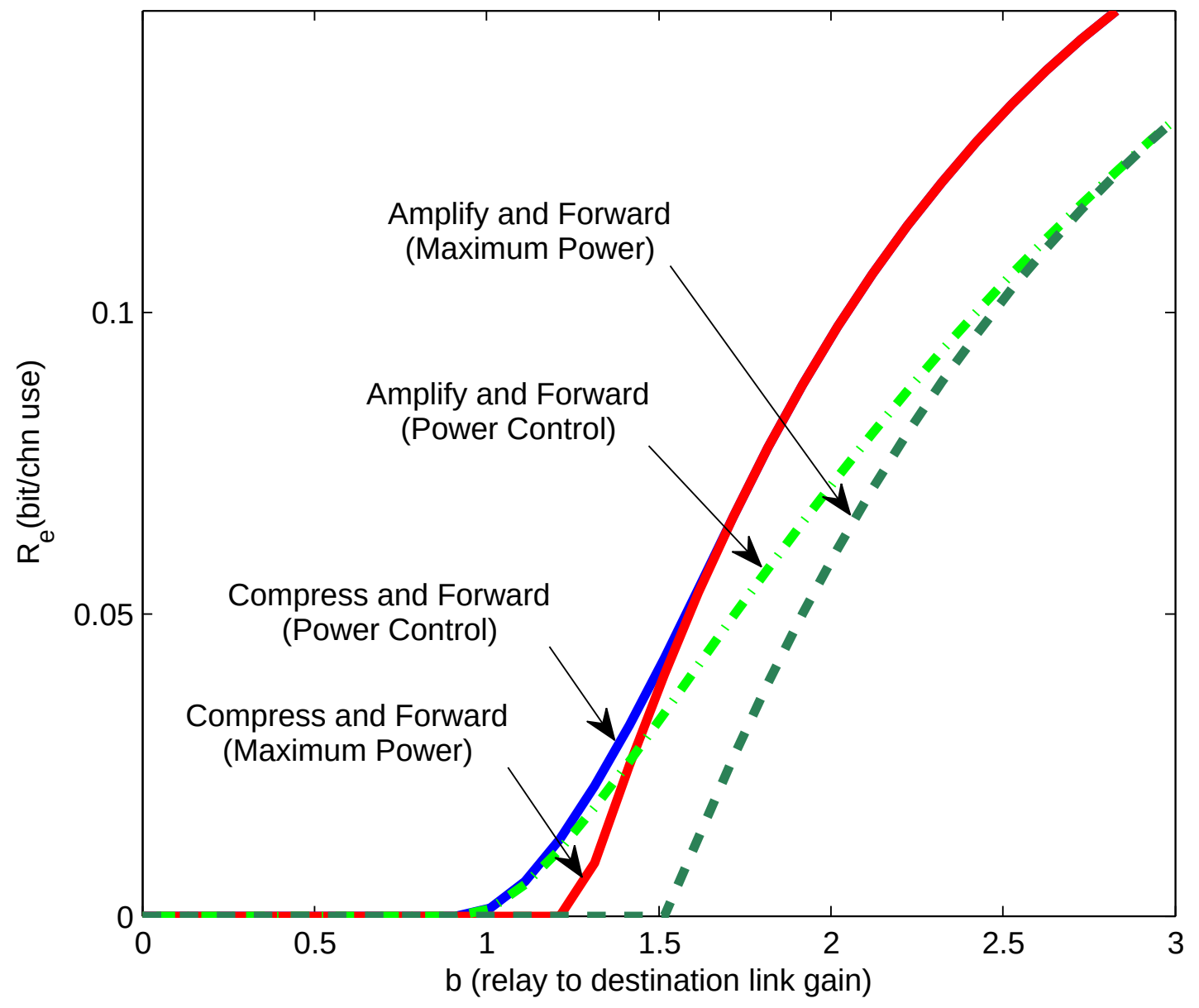

Fig. 7. Effect of Source Power Control

Observe that amplify-and-forward can also achieve a non-zero secrecy rate given a large enough $b$. However, comparing it to (31), we find that the secrecy rate given by amplify-and-forward is strictly smaller than the secrecy rate achievable by compress-and-forward. $\square$

*Remark 10:* When there are no secrecy constraints, for compress-and-forward/amplify-and-forward, the source should always transmit at maximum power. However, when there are secrecy constraints, for compress-and-forward/amplify-and-forward, the source may not transmit at maximum power. This can be shown as follows.

We first look at the case where there are no secrecy constraints. The rate for compress-and-forward follows from the maximum possible value of $R_1$ in Corollary 2, which is

$$R_1 = \max_{0 \le p \le P_1} C\left(p + \frac{a^2 p}{1 + \sigma_Q^2}\right) \tag{37}$$

where $\sigma_Q^2$ is given by (32). Recall that $p$ is the average transmission power of the source node. Hence we only need to show that

$$C\left(p + \frac{a^2 p}{1 + \sigma_Q^2}\right) \tag{38}$$

is a monotonic function of $p$ which is proved in Appendix C. Hence to maximize $R_1$ we should choose $p = P_1$.

The rate for amplify-and-forward is derived by ignoring the eavesdropper in Figure 6. The achievable rate is $I(X;Y_RY_D)$, which, using the Gaussian input distribution for $X$, equals

$$\max_{0\leq p\leq P_1} \frac{1}{2}\log\left(1+\left(1+\xi\right)p\right) \tag{39}$$

where $\xi$ is given by (36). To prove that (39) is maximized at $p = P_1$, it is sufficient to prove that $\xi p$ is a monotonically increasing function of $p$, which can be shown by rewriting $\xi p$ as:

$$\frac{a^2p}{1+\frac{1}{\beta^2b^2}} = \frac{a^2p}{1+\frac{a^2p+1}{P_rb^2}} = \frac{a^2}{\frac{1+\frac{1}{P_rb^2}}{p}+\frac{a^2}{P_rb^2}} \tag{40}$$

When there are secrecy constraints, the secrecy rate is not necessarily maximized at $p = P_1$. This can be observed in particular when $b$ (the relay to destination link gain) is small. In this case, for compress-and-forward, as shown in (32), the quantization noise $\sigma_Q^2$ will increase more rapidly with source power $p$. Similarly, for amplify-and-forward, the $\xi$ in (36) will decrease more rapidly with source power $p$. This, along with the negative term $-C(a^2p)$ present in (31) (35), may offset the benefits of having a larger source power $p$. This phenomenon is demonstrated numerically in Figure 7, where the source-to-relay channel gain $a = 1.2$. Both compress-and-forward and amplify-and-forward can achieve a larger secrecy rate when power control is used at the source. Moreover, compared to compress-and-forward, amplify-and-forward benefits more from judicious power allocation at the source. □

## VI. Upper Bound for the Secrecy Rate Of A Special Class of Model 2

### A. The Enhanced Channel

In this section, we describe the general methodology that we use to derive the upper bound. Our upper bound involves introducing a second eavesdropper. The focus of this section is to investigate the sufficient condition such that doing so will not decrease the secrecy capacity of the channel. In Section VI-B, this will be useful in finding the upper bound for the secrecy rate for a class of channels conforming to Model 2.

We focus on the case there is no feedback from the relay's output $X_r$ to its input $Y_r$, which means the conditional probability distribution of the channel should have the following form:

$$p(Y_r|X)p(Y|X,X_r,Y_r) \tag{41}$$

Note that due to the absence of feedback, we drop the term $X_r$ from the conditioning of $Y_r$. The reason that we choose this distribution to study will be clear shortly.

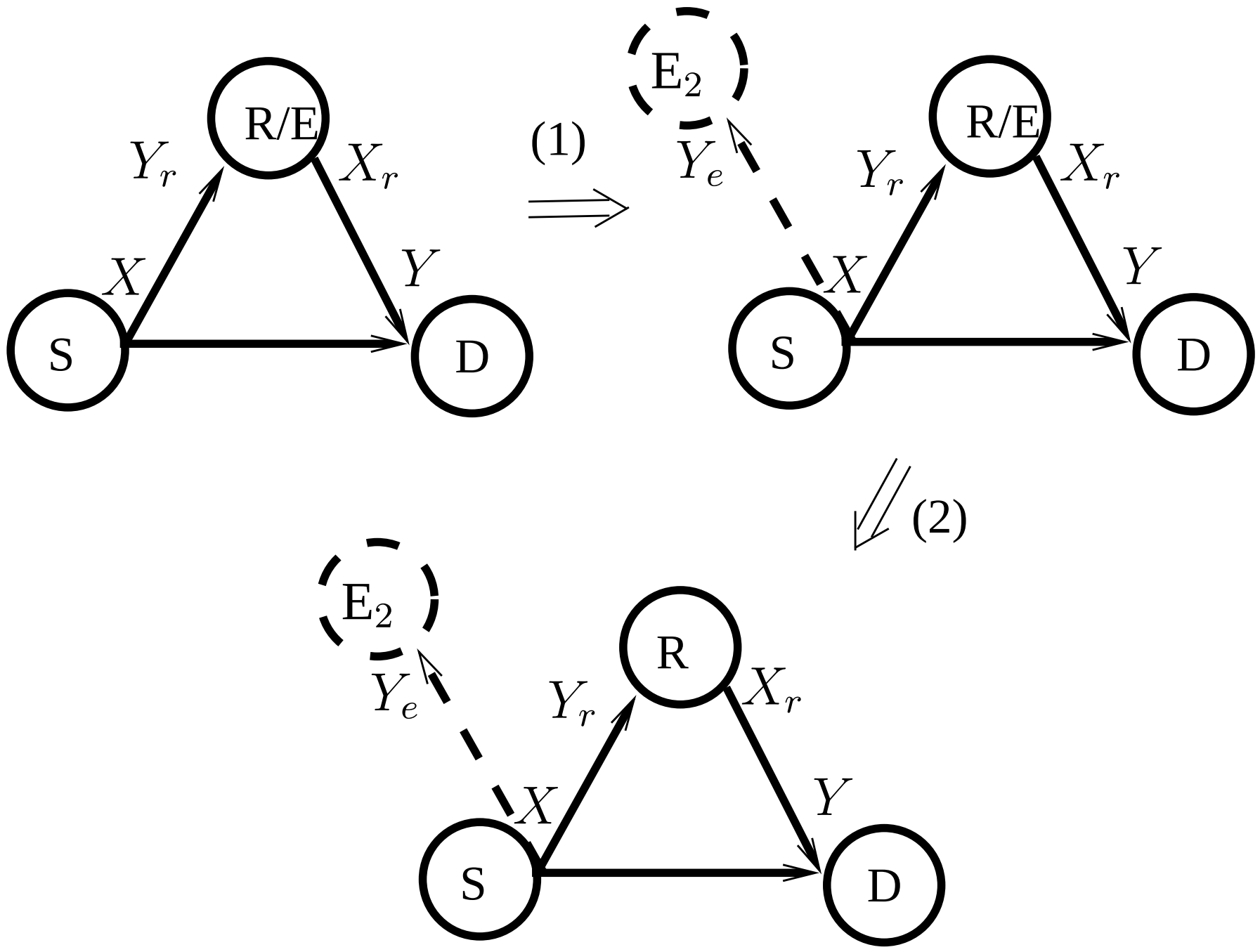

Fig. 8. The Relay-Eavesdropper Separation Argument

Deriving the outer bound entails a “relay-eavesdropper separation” argument. In other words, the net effect of this argument is to change the eavesdropper that is co-located with relay node, to an eavesdropper that is external to the relay node. Illustrated in Figure 8, this means:

1) We add a second eavesdropper to the relay network, who sees a channel that is statistically equivalent to the channel seen by the relay node. Let the signal received by this second eavesdropper be $Y_e$. That is, we have:

$$p(Y_r|X, X_r) = p(Y_e|X, X_r) \tag{42}$$

2) We remove the first eavesdropper.

The reader, at this point, rightfully should question the validity of step 1). This is because, as mentioned earlier, introducing a second eavesdropper, can *decrease* secrecy rate in general, even if the second eavesdropper observes a statistically equivalent channel as in (42). This is because the second eavesdropper may be able to *hear* the transmission signal $X_r$ of the first eavesdropper, and these two eavesdroppers can potentially cooperate. An example is provided in Appendix D to demonstrate this phenomenon.

We next show that, for the channel model in (41), introducing a second eavesdropper, if done with care, will not alter the secrecy capacity of the system. In particular, let the received signal of the "second" eavesdropper $Y_e$ be defined as follows:

$$p(Y_r|X)p(Y, Y_e|X, X_r, Y_r) \tag{43}$$

$$\text{s.t.} \quad p(Y_e|X, X_r) = p(Y_e|X) = p(Y_r|X) \tag{44}$$

Note that the second equality in (44) is (42) specialized for (41). We reiterate that though a $Y_e$ conforming to these conditions may not exist for any arbitrary relay network, for the Gaussian relay network models we are interested in, such a $Y_e$ can be found, as will be seen in the sequel (See (86), (110), (113)).

For this choice of $Y_e$, we have the following theorem:

*Theorem 4:* For the relay channel defined by (41), (43)-(44) are sufficient for the secrecy capacity of the channel after introducing the second eavesdropper to remain identical to the secrecy capacity of the original channel.

*Proof:* Due to the addition of the second eavesdropper, we know that the secrecy capacity of the new channel $\leq$ the secrecy capacity of the original channel. Therefore, we only need to show that the secrecy capacity of the new channel $\geq$ the secrecy capacity of the original channel.

We use $q$ to denote any distribution related to the new channel, and $p$ for any distribution related to the original channel. Suppose the new channel uses the exact same coding scheme and the same message set $\{W\}$ as the original channel. Then we can make the following statements:

1) Suppose $W$ can be reliably received by the destination at a rate of $R_e$ in the original channel. Then it must be reliably received by the receiver at the same rate in the new channel as well, because these two channels share the same coding scheme and the same channel statistics.
2) The transmitted message $W$ is still secret from the first eavesdropper co-located with the relay, since we are using the exact same coding scheme of the original channel.
3) We next show that $H(W|Y_e^n)$ of the new channel equals $H(W|Y_r^n)$ of the original channel. To do that, it is sufficient to prove that $q(Y_e^n|W)$ of the new channel equals $p(Y_r^n|W)$ of the original channel, as we show next:[2]

[2] It is understood in the case of continuous random variable, the sum should be replaced by integral. In fact, both of them can be expressed as integral by defining the measure properly.

First we state two Markov chains, which are proved in Appendix E.

$$\begin{aligned} &Y_{r,i} - X_i - X_{]i[}Y_{r,1}^{i-1} \\ &Y_{e,i} - X_i - X_{]i[}Y_{e,1}^{i-1} \end{aligned} \tag{45}$$

We have:

$$\begin{aligned} &q\left(Y_e^n|W\right) \\ =&\sum_{X^n} q\left(Y_e^n|X^n\right) q\left(X^n|W\right) &(46)\\ =&\sum_{X^n}\prod_{i=1}^{n} q\left(Y_{e,i}|X^n, Y_{e,1}^{i-1}\right) q\left(X^n|W\right) &(47)\\ \overset{(a)}{=}&\sum_{X^n}\prod_{i=1}^{n} q\left(Y_{e,i}|X_i\right) q\left(X^n|W\right) &(48)\\ \overset{(b)}{=}&\sum_{X^n}\prod_{i=1}^{n} p\left(Y_{r,i}|X_i\right) p\left(X^n|W\right) &(49)\\ \overset{(c)}{=}&\sum_{X^n}\prod_{i=1}^{n} p\left(Y_{r,i}|X^n, Y_{r,1}^{i-1}\right) p\left(X^n|W\right) &(50)\\ =&\sum_{X^n} p\left(Y_r^n|X^n\right) p\left(X^n|W\right) &(51)\\ =&p\left(Y_r^n|W\right) &(52) \end{aligned}$$

Here step $(a)$ follows from the Markov chain $Y_{e,i} - X_i - X_{]i[}Y_{e,1}^{i-1}$. Step $(b)$ follows from the fact that these two channels share the same coding scheme, $p(X^n|W) = q(X^n|W)$, and the constraint we placed on the marginal distribution $q(Y_e|X) = p(Y_r|X)$. Step $(c)$ follows the Markov chain $Y_{r,i} - X_i - X_{]i[}Y_{r,1}^{i-1}$.

The fact that introducing an eavesdropper does not reduce the secrecy capacity can then be seen from the following relationship:

$$\lim_{n\to\infty}\frac{1}{n}I\left(W;Y_r^n X_r^n\right) \geq \lim_{n\to\infty}\frac{1}{n}I\left(W;Y_r^n\right) = \lim_{n\to\infty}\frac{1}{n}I\left(W;Y_e^n\right) \geq 0 \tag{53}$$

If $\lim_{n\to\infty}\frac{1}{n}I\left(W;Y_r^n X_r^n\right) = 0$, then $\frac{1}{n}I\left(W;Y_e^n\right) = 0$. Therefore, for a given coding scheme, if $W$ is kept secret from the eavesdropper at the relay, it is also kept secret from the newly introduced eavesdropper. Hence any secrecy rate achievable in the original channel is achievable after introducing the second eavesdropper. This means the secrecy capacity remains the same. ∎

Theorem 4 shows that if the relay is not self-interfering, adding an eavesdropper as described in step 1 will not incur any loss in secrecy rate. This, along will step 2, will result in an "enhanced" channel whose secrecy rate is an upper bound to that of the original channel.

*Remark 11:* Actually, for the channel model in (41), we have

$$H(W|Y_r^n, X_r^n) = H(W|Y_r^n) \tag{54}$$

This means the secrecy capacity of the channel model in can be computed via $\lim_{n\to\infty} \frac{1}{n} H(W|Y_r^n)$ instead. This is proved in Appendix F.□

*Remark 12:* Note that the conditional probability distribution of the relay channel $p(Y, Y_r|X, X_r)$ is left *intact*. The benefit of the separation argument is that we have freedom in choosing $Y_e$, as long as it conforms to (43) and (44). Choosing $Y_e$ properly allows us to tighten the bound.□

### *B. Upper Bound for a Special Class of Model* 2

We next use the result we derived in Section VI-A to upper bound the secrecy rate of a class of relay channels. This class, as we mentioned earlier, is given by (16), which can be specialized from (41). Equation (43) becomes:

$$p(Y_D, Y_R, Y_r, Y_e|X, X_r) = p(Y_D|X)p(Y_r|X, Y_D)p(Y_R|X_r)p(Y_e|X, Y_D, Y_r) \tag{55}$$

$$p(Y_r|X) = p(Y_e|X) \tag{56}$$

*Definition 1:* Define $\mathcal{P}$ as the set of joint probability distribution functions of $Y_D, Y_R, Y_r, Y_e, X, X_r$ such that (55) and (56) are fulfilled.

With this definition, we have the following theorem:

*Theorem 5:* For the relay channel defined in (16), where the relay is the eavesdropper, the secrecy rate $R_e$ is upper bounded by

$$\max_{p(X,X_r)} \min \left\{ \begin{array}{l} I\left(X; Y_D|Y_r\right) \\ I\left(X_r; Y_R\right) + \min_{\mathcal{P}} I\left(X; Y_D|Y_e\right) \end{array} \right\} \tag{57}$$

*Proof:* The first term can be obtained by specializing the result from [21]. Reference [21, version 7,(13)] claims for a general relay $p(Y, Y_r|X, X_r)$, the secrecy rate is upper bounded by

$$R_e \leq I(X; Y|Y_r, X_r) \tag{58}$$

Specializing it to our channel, which means replacing $Y$ with $Y_D, Y_R$, we have

$$I\left(X;Y_D,Y_R|Y_r,X_r\right) \tag{59}$$

$$\leq I\left(X;Y_R|Y_r,X_r\right)+I\left(X;Y_D|Y_r,X_r,Y_R\right) \tag{60}$$

From (55), $X-\{Y_r,X_r\}-Y_R$ is a Markov chain. Hence (60) equals:

$$I\left(X;Y_D|Y_r,X_r,Y_R\right) \tag{61}$$

$$=h\left(Y_D|Y_r,X_r,Y_R\right)-h\left(Y_D|Y_r,X_r,X,Y_R\right) \tag{62}$$

$$\leq h\left(Y_D|Y_r\right)-h\left(Y_D|Y_r,X_r,X,Y_R\right) \tag{63}$$

From (55), $Y_D-\{X,Y_r\}-\{X_r,Y_R\}$ is a Markov chain. Hence (63) equals

$$h\left(Y_D|Y_r\right)-h\left(Y_D|X,Y_r\right)=I\left(X;Y_D|Y_r\right) \tag{64}$$

Hence we have proved the first term.

Next, we proceed to bound the second term:

$$H\left(W|Y_e^n\right)$$

$$\overset{(a)}{\leq} I\left(W;Y_D^nY_R^n|Y_e^n\right)+n\varepsilon_1 \tag{65}$$

$$=I\left(W;Y_D^n|Y_e^n\right)+I\left(W;Y_R^n|Y_e^nY_D^n\right)+n\varepsilon_1 \tag{66}$$

$$\leq I\left(WX^n;Y_D^n|Y_e^n\right)+I\left(WX_r^n;Y_R^n|Y_e^nY_D^n\right)+n\varepsilon_1 \tag{67}$$

$$=I\left(X^n;Y_D^n|Y_e^n\right)+I\left(WX_r^n;Y_R^n|Y_e^nY_D^n\right)+n\varepsilon_1 \tag{68}$$

$$=I\left(X^n;Y_D^n|Y_e^n\right)+h\left(Y_R^n|Y_e^nY_D^n\right)-\sum_{i=1}^{n}h\left(Y_{R,i}|Y_e^nY_D^nX_r^nY_{R,1}^{i-1}W\right)+n\varepsilon_1 \tag{69}$$

$$\overset{(b)}{=} I\left(X^n;Y_D^n|Y_e^n\right)+h\left(Y_R^n|Y_e^nY_D^n\right)-\sum_{i=1}^{n}h\left(Y_{R,i}|X_{r,i}\right)+n\varepsilon_1 \tag{70}$$

$$\leq I\left(X^n;Y_D^n|Y_e^n\right)+\sum_{i=1}^{n}\left(h\left(Y_{R,i}\right)-h\left(Y_{R,i}|X_{r,i}\right)\right)+n\varepsilon_1 \tag{71}$$

$$=h\left(Y_D^n|Y_e^n\right)-\sum_{i=1}^{n}h\left(Y_{D,i}|Y_e^nX^nY_{D,1}^{i-1}\right)+\sum_{i=1}^{n}I\left(X_{r,i};Y_{R,i}\right)+n\varepsilon_1 \tag{72}$$

$$\overset{(c)}{=} h\left(Y_D^n|Y_e^n\right)-\sum_{i=1}^{n}h\left(Y_{D,i}|Y_{e,i}X_i\right)+\sum_{i=1}^{n}I\left(X_{r,i};Y_{R,i}\right)+n\varepsilon_1 \tag{73}$$

$$\leq \sum_{i=1}^{n}I\left(X_i;Y_{D,i}|Y_{e,i}\right)+\sum_{i=1}^{n}I\left(X_{r,i};Y_{R,i}\right)+n\varepsilon_1 \tag{74}$$

$$=nI\left(X;Y_D|Y_e,Q\right)+I\left(X_r;Y_R|Q\right)+n\varepsilon_1 \quad (75)$$

$$=nh\left(Y_D|Y_e,Q\right)-nh\left(Y_D|Y_e,X\right)+nh\left(Y_R|Q\right)-nh\left(Y_R|X_r\right)+n\varepsilon_1 \quad (76)$$

$$\leq nh\left(Y_D|Y_e\right)-nh\left(Y_D|Y_e,X\right)+nh\left(Y_R\right)-nh\left(Y_R|X_r\right)+n\varepsilon_1 \quad (77)$$

$$=nI\left(X;Y_D|Y_e\right)+nI\left(X_r;Y_R\right)+n\varepsilon_1 \quad (78)$$

Here step $(a)$ follows from Fano's inequality. Step $(b)$ follows from the relay destination link being orthogonal to the rest part of the channel. Step $(c)$ follows from the fact that the relay is not interfering the second eavesdropper. Therefore given $\{Y_{e,i}, X_i\}$, the signals $\{Y_{e,j}, j > i\}$ do not provide further information about $Y_{D,i}$. ■

*Remark 13:* Another upper bound that can be obtained is

$$I\left(X;Y_DY_r|Y_e\right) \quad (79)$$

which is proved in Appendix G, and it can be further tightened by choosing $Y_e$. However, as shown below this upper bound does not improve the first term in (57).

$$I\left(X;Y_DY_r|Y_e\right)$$

$$=I\left(X;Y_DY_rY_e\right)-I\left(X;Y_e\right) \quad (80)$$

$$\overset{(b)}{=}I\left(X;Y_DY_rY_e\right)-I\left(X;Y_r\right) \quad (81)$$

$$\geq I\left(X;Y_DY_r\right)-I\left(X;Y_r\right) \quad (82)$$

$$=I\left(X;Y_D|Y_r\right) \quad (83)$$

Step $(b)$ follows from $p(Y_r|X)=p(Y_e|X)$. □

### *C. The Gaussian Case of Model* 2

Using Theorem 5, we now evaluate the upper bound for the Gaussian channel.

*Corollary 3:* For the Gaussian case of Model 2, which has independent noise components, the upper bound on secrecy rate is:

$$\min\left\{C(b^2P)+[C(P)-C(a^2P)]^+, C\left(\frac{P}{1+a^2P}\right)\right\} \quad (84)$$

*Proof:* First we notice that (57) is upper bounded by:

$$\min\left\{\begin{array}{l}\max\limits_{p(X,X_r)} I\left(X;Y_D|Y_r\right)\\ \max\limits_{p(X_r)} I\left(X_r;Y_R\right)+\min\limits_{\mathcal{P}}\max\limits_{p(X,X_r)} I\left(X;Y_D|Y_e\right)\end{array}\right\} \quad (85)$$

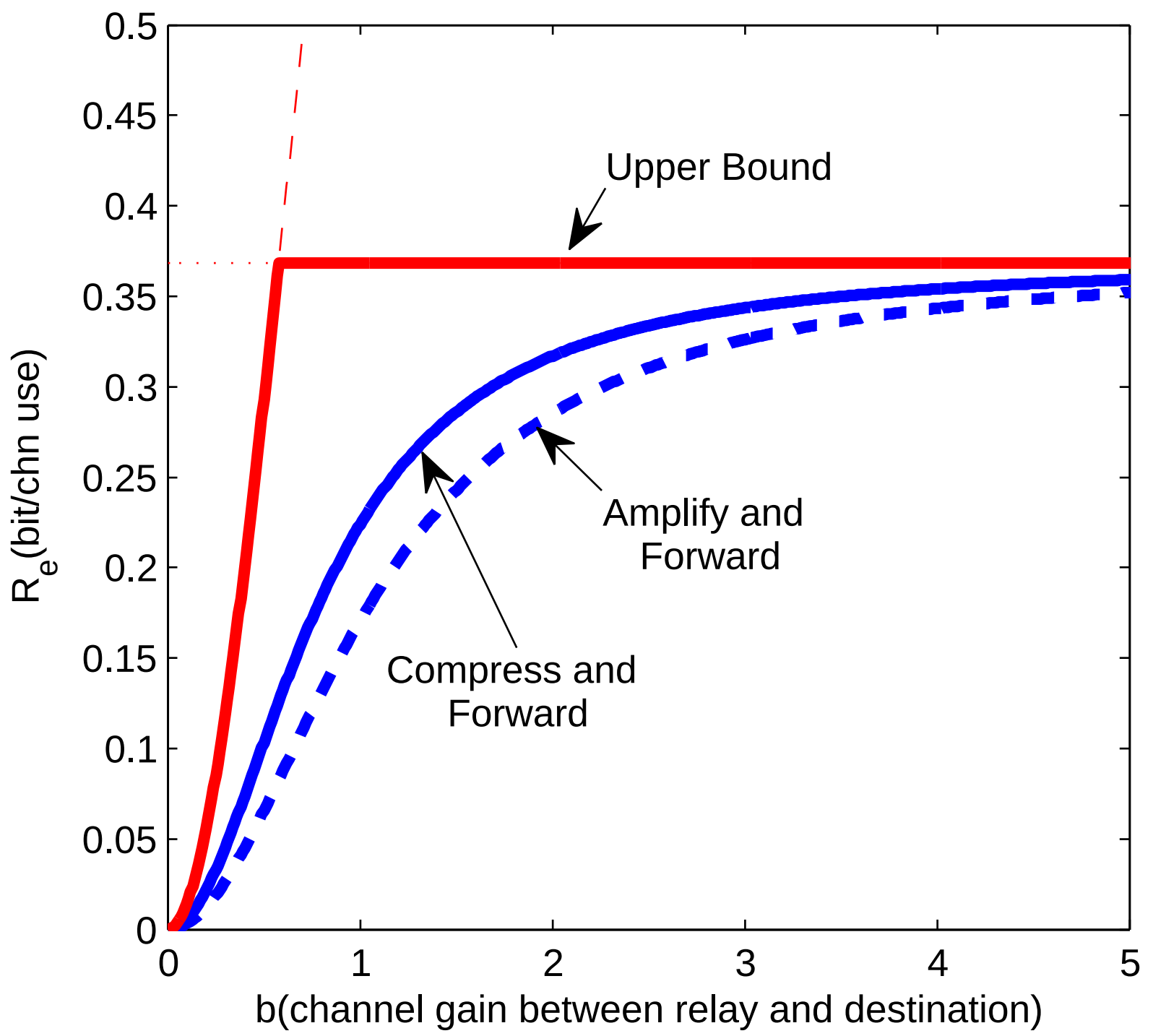

Fig. 9. Secrecy Rate of the Gaussian Orthogonal Relay Channel

Since $p(Y_e|X) = p(Y_r|X)$, we define a Gaussian random variable $N_e$ such that

$$Y_e = aX + N_e \tag{86}$$

Then the set $\mathcal{P}$ can be re-parametrized with the correlation between $N_e, N_r$ and the correlation between $N_e, N_D$. For a given correlation, it is known that $I(X; Y_D|Y_e)$ and $I(X; Y_rY_D|Y_e)$ are both maximized with Gaussian distribution [5, Appendix II]. The first term in (85) then becomes $\frac{1}{2}\log_2\left(1 + \frac{P}{1+a^2P}\right)$. To obtain the second term inside the minimum, we choose $N_e/a^2 = N_D + N'$, with $N', N_D$ being independent if $a < 1$, otherwise we choose $N_D = N_e/a^2 + N'$, with $N', N_e$ being independent, where $N'$ is a zero mean Gaussian random variable with appropriate variance.

■

In Figure 9, we compare the upper bound with the achievable rates for the Gaussian case of Model 2. We fix the source-to-relay channel gain $a = 1$, and vary the relay-to-destination channel gain $b$. As $b \to \infty$, we observe that the upper bound becomes tight. As $b \to 0$, the upper bound decreases. This improvement is due to the first term in Corollary 3.

## VII. THE COVER-KIM DETERMINISTIC RELAY CHANNEL

In this section, we investigate the Cover-Kim deterministic relay channel of [24], [28] whose capacity is established therein. The channel was defined in Figure 5 in Section III.

For the achievable secrecy rate, we have the next theorem.

*Theorem 6:* For the Gaussian Cover-Kim deterministic channel, the following secrecy rate is achievable:

$$\left[R_0 + C\left(P\right) - C\left(\alpha^2 P\right)\right]^+ \tag{87}$$

*Proof:* Let $\mathcal{C}$ be a random code book with $2^{\lfloor n\left[R_0+C(P)-C\left(\alpha^2 P\right)\right]^+\rfloor}2^{\lfloor nC\left(\alpha^2 P\right)\rfloor}$ codewords sampled from an i.i.d. Gaussian distribution with zero mean and variance $P$. These codewords are randomly partitioned into $2^{\lfloor n\left[R_0+C(P)-C\left(\alpha^2 P\right)\right]^+\rfloor}$ bins of equal size. The bin index of the transmitted codeword is determined by the message $W$. The actual transmitted codeword is then selected randomly from this bin according to a uniform distribution. The relay uses either hash-and-forward or compress-and-forward as described in [24]. Let $E[P_e|\mathcal{C}]$ be the average error probability over the codebook ensemble $\{\mathcal{C}\}$ that the destination could not correctly determine $X^n$, hence $W$, from $Y_D^n$ and side information provided by the relay. It was proved in [24] that $\lim_{n\to\infty} E[P_e|\mathcal{C}] = 0$.

Since each bin is a Gaussian codebook by itself whose rate is below the AWGN channel capacity between the source and the relay, the relay node can determine $X^n$ given $W$ and $Y_r^n$ with high probability using jointly typical decoding. Therefore, from Fano's inequality, we have $H\left(X^n|WY_r^n\mathcal{C}\right) \leq n\varepsilon_1$. Thus:

$$H\left(W|Y_r^n\mathcal{C}\right) = H\left(X^nW|Y_r^n\mathcal{C}\right) - H\left(X^n|WY_r^n\mathcal{C}\right) \tag{88}$$

$$\geq H\left(X^nW|Y_r^n\mathcal{C}\right) - n\varepsilon_1 \tag{89}$$

$$= H\left(X^n|Y_r^n\mathcal{C}\right) + H\left(W|X^nY_r^n\mathcal{C}\right) - n\varepsilon_1 \tag{90}$$

$$= H\left(X^n|Y_r^n\mathcal{C}\right) - n\varepsilon_1 \tag{91}$$

$$= H\left(X^n|\mathcal{C}\right) - I\left(X^n;Y_r^n|\mathcal{C}\right) - n\varepsilon_1 \tag{92}$$

$$\geq H\left(X^n|\mathcal{C}\right) - I\left(X^n;Y_r^n\right) - n\varepsilon_1 \tag{93}$$

$$\geq H\left(X^n|\mathcal{C}\right) - \sum_{i=1}^{n} I\left(X_i;Y_{r,i}\right) - n\varepsilon_1 \tag{94}$$

Since each code word is selected with equal probability, we have

$$\lim_{n\to\infty}\frac{1}{n}H\left(X^n|\mathcal{C}\right)=C(P)+R_0 \tag{95}$$

Also, $I\left(X_i;Y_{r,i}\right)=C(\alpha^2P)$. Substituting this and (95) into (94), dividing it by $n$ and taking the limit $n\to\infty$, we have (87), which equals $\lim_{n\to\infty}\frac{1}{n}H(W|\mathcal{C})$. Therefore $\lim_{n\to\infty}E[P_e|\mathcal{C}]+\frac{1}{n}I(W;Y_r^n|\mathcal{C})=0$. Since both terms inside the limit are non-negative, this proves the existence of at least one codebook with a rate of $\left[R_0+C\left(P\right)-C\left(\alpha^2P\right)\right]^+$ such that both terms are arbitrarily small. Hence we have proved the theorem. ■

*Theorem 7:* The secrecy rate of the Gaussian Cover-Kim deterministic channel is upper bounded by

$$R_0+\left[C\left(P\right)-C\left(\alpha^2P\right)\right]^+ \tag{96}$$

*Proof:* We use Theorem 4 to separate the eavesdropper and the relay. Let $Y_e$ be the signal received by the eavesdropper such that (55) and (56) are met. Then, we have:

$$\begin{aligned}
&H\left(W|Y_e^n\right)\\
&\overset{(a)}{\leq} I\left(W;Y_D^nX_R^n|Y_e^n\right)+n\varepsilon_1 && (97)\\
&=I\left(W;Y_D^n|Y_e^n\right)+I\left(W;X_R^n|Y_e^nY_D^n\right)+n\varepsilon_1 && (98)\\
&\leq I\left(W;Y_D^n|Y_e^n\right)+H\left(X_R^n\right)+n\varepsilon_1 && (99)\\
&\leq I\left(WX^n;Y_D^n|Y_e^n\right)+H\left(X_R^n\right)+n\varepsilon_1 && (100)\\
&=I\left(X^n;Y_D^n|Y_e^n\right)+H\left(X_R^n\right)+n\varepsilon_1 && (101)\\
&=h\left(Y_D^n|Y_e^n\right)-\sum_{i=1}^{n}h\left(Y_{D,i}|Y_e^nX^nY_{D,1}^{i-1}\right)+H\left(X_R^n\right)+n\varepsilon_1 && (102)\\
&\overset{(b)}{=}h\left(Y_D^n|Y_e^n\right)-\sum_{i=1}^{n}h\left(Y_{D,i}|Y_{e,i}X_i\right)+H\left(X_R^n\right)+n\varepsilon_1 && (103)\\
&\leq\sum_{i=1}^{n}I\left(X_i;Y_{D,i}|Y_{e,i}\right)+H\left(X_R^n\right)+n\varepsilon_1 && (104)\\
&=nI\left(X;Y_D|Y_e,Q\right)+H\left(X_R^n\right)+n\varepsilon_1 && (105)\\
&=nh\left(Y_D|Y_e,Q\right)-nh\left(Y_D|Y_e,X\right)+H\left(X_R^n\right)+n\varepsilon_1 && (106)\\
&\leq nh\left(Y_D|Y_e\right)-nh\left(Y_D|Y_e,X\right)+H\left(X_R^n\right)+n\varepsilon_1 && (107)\\
&=nI\left(X;Y_D|Y_e\right)+H\left(X_R^n\right)+n\varepsilon_1 && (108)
\end{aligned}$$

$$\leq nI\left(X;Y_D|Y_e\right)+nR_0+n\varepsilon_2 \tag{109}$$

Step $(a)$ follows from Fano's inequality. Step $(b)$ follows from the fact that the relay is not interfering with, i.e., heard by the (second) eavesdropper. Therefore, given $\{Y_{e,i},X_i\}$, signals $\{Y_{e,j},j>i\}$ will not provide more information about $Y_{D,i}$.

The bound is further tightened by choosing $Y_e$ properly.

1) If $\alpha\geq 1$, then

$$Y_e=\alpha X+Z \tag{110}$$

$$Y_D=X+\frac{Z}{\alpha}+Z' \tag{111}$$

$$Y_r=X-\frac{Z}{\alpha}-Z' \tag{112}$$

   $Z'$ is a zero mean Gaussian random variable with variance $|1-\frac{1}{\alpha^2}|$, and $Z'$ is independent from $Z$.

2) If $\alpha\leq 1$, then

$$Y_e=X+Z+Z' \tag{113}$$

   $Z'$ is a zero mean Gaussian random variable with variance $|1-\frac{1}{\alpha^2}|$, and $Z'$ is independent from $Z$.

Substituting these choices of $Y_e$ into (109), we get (96). ■

*Remark 14:* Inspecting (87) and (96), we see that the upper bound and the achievable rate coincide when $\alpha\leq 1$. Hence, for $\alpha\leq 1$, i.e, when the source to destination link is not worse than the source to the relay link, the secrecy capacity is achieved by compress-and-forward. □

*Remark 15:* The secrecy capacity can exceed the direct link capacity if $R_0>C(P)$. This is a benefit of the correlation of the noises corrupting the links from the source. If the noises are independent, the secrecy capacity cannot exceed $C(P)$, as proved next:

*Observation 1:* If the relay channel has the property:

$$p\left(Y_R,Y_D,Y_r|X,X_r\right)=p\left(Y_R|X_r\right)p\left(Y_r|X\right)p\left(Y_D|X\right) \tag{114}$$

Then $R_e\leq I\left(X;Y_D\right)$

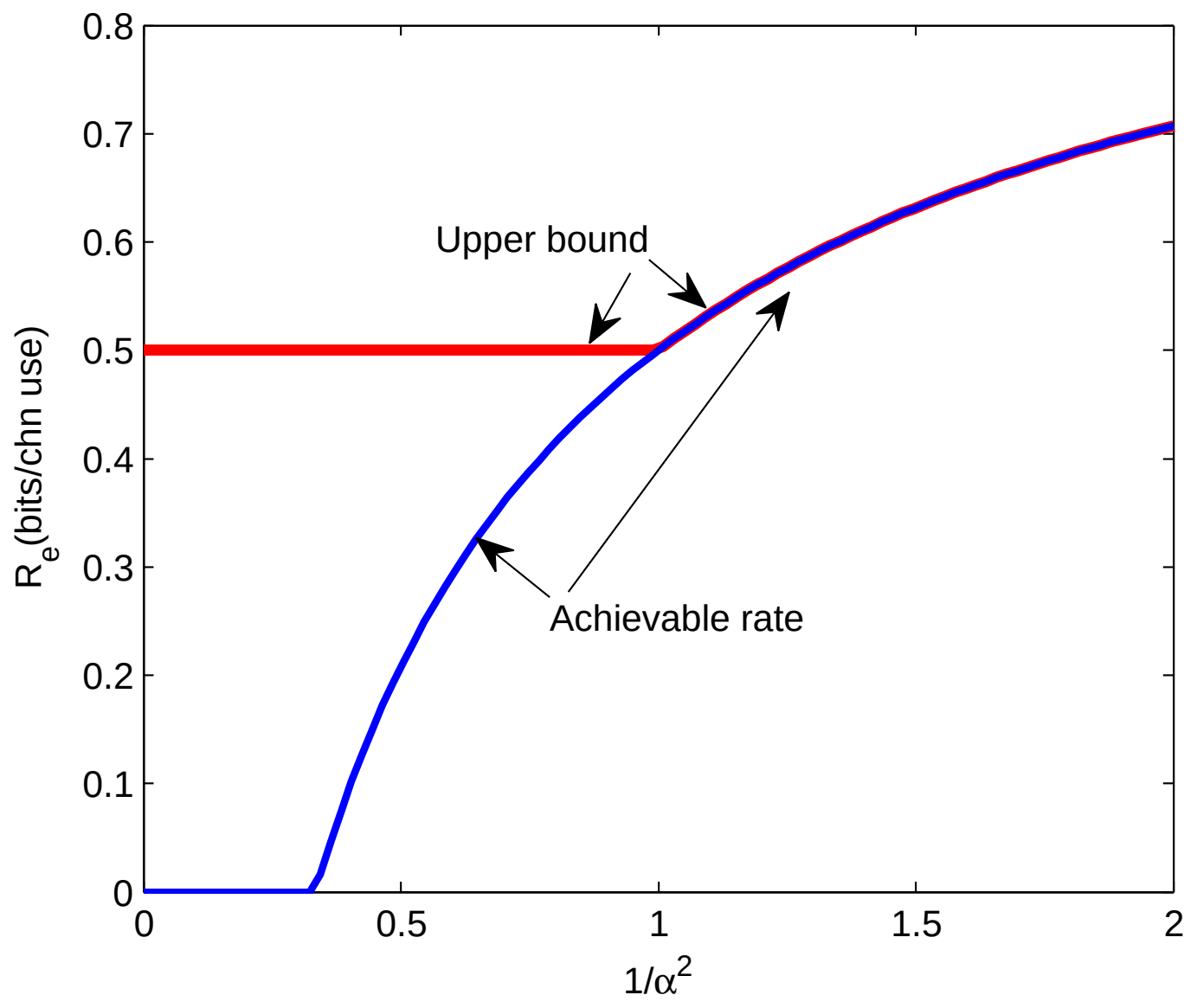

Fig. 10. Secrecy rate for the Gaussian Cover-Kim Deterministic Relay channel

*Proof:* From (57), we have

$$R_e \leq I\left(X;Y_D|X_r\right) \tag{115}$$

$$= h\left(Y_D|X_r\right) - h\left(Y_D|X_r, X\right) \tag{116}$$

$$= h\left(Y_D|Y_r\right) - h\left(Y_D|X\right) \tag{117}$$

$$\leq I\left(X;Y_D\right) \tag{118}$$

where in (117) we use the Markov chain $Y_D - X - Y_r$. ■

□

We conclude this section by presenting Figure 10 which shows the upper bound and the achievable rate for $R_0 = 0.5$ bits/channel use and $P = 1$. As expected, the two meet for $\alpha \leq 1$, yielding the secrecy capacity.

## VIII. Conclusion

In this paper, we have considered the relay channel with an untrusted relay that is treated as an eavesdropper. In particular, we focused on two relay channel models with orthogonal components. For the first model, we have found the capacity-equivocation region and proved that the relay-destination link does not help in increasing secrecy rate, and therefore the untrusted relay should not be deployed if perfect secrecy is desired. In contrast, for the second model, we have found an achievable secrecy rate which calls relay's cooperation and improves the secrecy rate as compared to treating it simply as an eavesdropping node. Thus, we conclude that, for this model, the untrusted relay may help the source and the destination to communicate despite being subjected to the secrecy constraint, and that cooperation with the untrusted relay is beneficial.

We have provided a channel transformation that separates the relay and the eavesdropper to upper bound the secrecy rate for a special class of untrusted relay channels. We have found this approach to be useful in upper bounding the secrecy rate for two cases: For the Gaussian relay channel with an orthogonal relay-destination link, this new approach yields a computable bound that tightens previously known bounds. For the Gaussian Cover-Kim deterministic relay channel, we have shown that this approach finds the secrecy capacity when the source-destination link is not worse than the source-relay link.

Since the first example demonstrating the potential benefit of cooperating with an untrusted relay [29], there has been recent growing interest in communication models with untrusted relays. Notable recent developments include work on the multiple access channel with generalized feedback [30] and relay broadcast channel [31], [32], where, in addition to the secret message considered in this work, the untrusted relay node has its own secret message. The role of untrusted relay is examined in bi-directional communication in [33], [34], where the relay node in a two-way relay network is untrusted. A case for the communication scenario with multiple untrusted relay nodes is recently presented in [35], [36], where the source and the destination can only communicate via a chain of untrusted relay nodes. All these works, like this paper we are about to conclude, speaks to the merit of cooperative communication even with untrusted partners, and that cooperation and secrecy can go hand in hand.

## APPENDIX A
## PROOF OF THEOREM 1

The achievable scheme of Theorem 1 is a combination of stochastic encoding at the source node and compress-and-forward at the relay node. The compress-and-forward relaying scheme is the same one described in [25]. The achievable scheme involves $N$ blocks of channel uses. Each block is composed of $n$ channel uses.

### A. Codebook Generation

*1) The Codebook of the Source Node:* The source uses a codebook composed of i.i.d. sequences sampled from the distribution $p(X)$. Each codeword has $n$ components. In order to confuse the relay/eavesdropper, the codebook is further partitioned randomly to bins. Suppose there are $2^{nC}$ bins. Each bin contains $2^{nB}$ codewords. $B$ is chosen such that:

$$nB = \lfloor nI(X; Y_r|X_r) \rfloor \tag{119}$$

The reason behind this choice will be clear shortly. Each codeword is hence indexed by the label $\{b, c\}$, where $c$ is the bin index and $b$ indexes the codeword within the bin. The rate of the codebook is given by

$$\lim_{n\to\infty} \frac{1}{n} \log_2 |X^n| = \lim_{n\to\infty} B + C = I\left(X; Y_r|X_r\right) + \lim_{n\to\infty} C \tag{120}$$

*2) The Codebook of the Relay [25]:*

i) The signal transmitted by the relay is from a codebook composed of i.i.d. sequences sampled from the distribution $p(X_r)$. Each codeword has $n$ components and is denoted by $X_r^n(d)$. The codebook has $2^{nD}$ codewords.

ii) For each $d$, we generate $2^{nE}$ codewords, each with $n$ components, denoted by $\hat{Y}_r(e|d)$. The $i$th component of the codeword is drawn from $p(\hat{Y}_r|X_r = X_{r,i}^n(d))$ in an i.i.d. fashion.

iii) For each $d$, we randomly bin the label $e$ into $2^{nD}$ bins and label each bin with a $d$ according to uniform distribution. This random binning is used for Wyner-Ziv coding.

We use $\mathcal{C}$ to denote the random codebooks generated for the source and the relay.

### B. Stochastic Encoder at the Source Node

The codeword transmitted as the $k$th block is indexed by label $b_k, c_k$, where $c_k$ is the bin index and $b_k$ indexes the codeword within the bin. Let $W(k)$ be the message transmitted at the $k$th block. Recall that $R_1$ is the rate of the message $W(k)$. Hence $R_1 = \log_2 |W(k)|/n$. The messages are mapped to the codewords as follows.

i) If $R_1 > C$, $c_k$ is the bin index determined by $W(k)$. The codewords in bin $c_k$ are partitioned into $2^{n(R_1-C)}$ subsets. The subset is chosen according to the unmapped part of $W(k)$. Then $b_k$ is selected from this chosen subset according to a uniform distribution.

ii) If $R_1 \leq C$, $c_k$ is still determined by $W(k)$. $b_k$ is randomly chosen from group $c_k$ according to a uniform distribution.

For this mapping, we observe that the cardinality of $c_k$ is $2^{n\min\{R_1,C\}}$.

Only $N-1$ messages $W(1)...W(N-1)$ are transmitted over $N$ blocks. During the last block, the relay and the source agrees that the source will send message $1$.

### *C. Compress-and-forward at the relay [25]*

During the $k$th block, the relay node first compresses $Y_r^n(k)$ to $\hat{Y}_r^n(e_k|d_k)$. $\hat{Y}_r^n$ is indexed by two labels: $e_k$ and $d_k$. $d_k$ is chosen to be the label that corresponds to $X_r^n(k)$. Hence a different set of $\hat{Y}_r^n$, of size $2^{nE}$, is used for compression depending on the value of $X_r^n(k)$. The label $e_k$ is chosen to be the first element in the following set:

$$\left\{e : \hat{Y}_r^n\left(e|d_k\right), Y_r^n\left(k\right), X_r^n\left(k\right) \text{ are jointly typical}\right\} \tag{121}$$

If the set is empty, $e_k = 1$. The size of the codebook $\hat{Y}_r^n(e|d_k)$ should be sufficiently large for the set to be nonempty, which requires $E > I(\hat{Y}_r; Y_r|X_r)$ [25].

Label $e_k$ is transmitted during the $k+1$st block. At this time the destination has received $Y^n(k)$, and can decode $X_r^n(k)$. Since $\{Y^n(k), X_r^n(k)\}$ provide side information to the destination about $e_k$, $e_k$ can be compressed further before transmission. This is done via Wyner-Ziv coding. Recall that the set $\{e, d = d_k\}$ is randomly binned. The size of each bin should be chosen such that the destination can decode $e_k$, and hence determine $\hat{Y}_r^n(k)$ from this bin from the side information $\{Y^n(k), X_r^n(k)\}$. This requires $I(\hat{Y}_r; Y|X_r) > E - D$. Only the bin index is transmitted. Recall that each bin is labeled with $d$. Hence this determines $d_{k+1}$, which determines $X_r^n(k+1)$.

*Remark 16:* One important aspect of this coding scheme is that the signals transmitted by the relay during different blocks $\{X_r^n(k), k = 1...N-1\}$ are correlated. This is because, as described in the coding scheme at the relay, each $X_r^n(k)$ is determined from $\hat{Y}_r^n(e_{k-1}|d_{k-1})$, which is shown in (121) to be related to $X_r^n(k-1)$. Because of the self interference at the relay, the signals received by the relay during different blocks $\{Y_r^n(k), k = 1...N-1\}$ are correlated as well. However, $Y_r^n(k)$ is correlated with past $Y_r^n(p), p < k$ only through $X_r^n(k)$. This property will be useful in bounding the equivocation rate. $\square$

### D. Decoder at the Destination

Recall that the short hand $W(a, ..., b)$ stands for the set $\{W(a), W(a+1), ..., W(b)\}$. The short hand $W^a$ stands for the set $\{W(1), W(2), ..., W(a)\}$.

The destination first decodes $X^{n(N-1)}$. The decoding at the $k$th block happens as: It first decodes $X_r^n(k)$ from $Y^n(k)$. For this, we require $D < I(X_r; Y)$. It then determines $d_k$ from $X_r^n(k)$, which determines the bin that contains $e_{k-1}$. It next determines $e_{k-1}$ by finding the label $e$ in this bin such that $\hat{Y}_r^n(e|d_{k-1})$ is joint typical with $\{Y^n(k-1), X_r^n(k-1)\}$. This determines $\hat{Y}_r^n(e_{k-1}|d_{k-1})$. Finally $X^n(k-1)$ is decoded from $Y^n(k-1), \hat{Y}_r^n(k-1), X_r^n(k-1)$. For details, the reader is referred to [25, Theorem 6].

Let the decoding result be $\hat{X}^{n(N-1)}$. According to error probability analysis in [25], if the rate of the codebook of the source meets the condition:

$$\lim_{n\to\infty} \frac{1}{n} \log_2 |X^n| < I(X; \hat{Y}_r, Y|X_r) \tag{122}$$

and the following condition is fulfilled:

$$I(X_r; Y) > I(\hat{Y}_r; Y_r|Y X_r) \tag{123}$$

then

$$\lim_{n\to\infty} E[\Pr(\hat{X}^{n(N-1)} \neq X^{n(N-1)}|\mathcal{C})] = 0 \tag{124}$$

The expectation is taken over the random codebook $\mathcal{C}$.

Combining (122) and (120), we have:

$$I(X; Y_r|X_r) + \lim_{n\to\infty} C < I(X; \hat{Y}_r, Y|X_r) \tag{125}$$

The destination then computes $\hat{W}^{N-1}$ from $\hat{X}^{n(N-1)}$, since the former is a deterministic function of the latter. The average probability of decoding error for $W^{n(N-1)}$ is hence upper bounded by the average probability of decoding error of $X^{n(N-1)}$. Therefore equation (124) implies:

$$\lim_{n\to\infty} E[\Pr(\hat{W}^{N-1} \neq W^{N-1}|\mathcal{C})] = 0 \tag{126}$$

### E. Equivocation Computation

Let $c^{N-1}$ denote $\{c_1, c_2, ..., c_{N-1}\}$. The computation of the equivocation rate starts from the following expression:

$$H(c^{N-1}|Y_r^{nN}, X_r^{nN}, \mathcal{C}) \tag{127}$$

$$=H(c^{N-1}|Y_r^{n(N-1)}, X_r^{nN}, \mathcal{C}) \tag{128}$$

$$=H(c^{N-1}|Y_r^{n(N-1)}, X_r^{n(N-1)}, \mathcal{C}) \tag{129}$$

Here, the first equality follows from the fact that conditioned on $X_r^n(N)$, the signals $Y_r^n(N)$ are independent from $Y_r^{n(N-1)}$, $X_r^{n(N-1)}$, $c^{N-1}$. The second equality is because, as described in the relaying scheme above, $X_r^n(N)$ is a deterministic function of $Y_r^n(N-1)$ and $X_r^n(N-1)$.

To simplify the notation, we omit the $\mathcal{C}$ from the conditioning term in the derivation below and only mention it when necessary.

Equation (129) can be reformulated as:

$$H\left(c^{N-1}|Y_r^{(N-1)n}, X_r^{(N-1)n}\right) \tag{130}$$

$$=-H\left(X^{(N-1)n}|c^{N-1}, Y_r^{(N-1)n}, X_r^{(N-1)n}\right) + H\left(c^{N-1}|X^{(N-1)n}, Y_r^{(N-1)n}, X_r^{(N-1)n}\right)$$
$$+H\left(X^{(N-1)n}|Y_r^{(N-1)n}, X_r^{(N-1)n}\right) \tag{131}$$

$$=-H\left(X^{(N-1)n}|c^{N-1}, Y_r^{(N-1)n}, X_r^{(N-1)n}\right) + H\left(X^{(N-1)n}|Y_r^{(N-1)n}, X_r^{(N-1)n}\right) \tag{132}$$

$$=-H\left(X^{(N-1)n}|c^{N-1}, Y_r^{(N-1)n}, X_r^{(N-1)n}\right) + H\left(X^{(N-1)n}\right)$$
$$-I\left(X^{(N-1)n}; Y_r^{(N-1)n}, X_r^{(N-1)n}\right) \tag{133}$$

From the description of the stochastic encoder at the source node, we observe that each block $X^n(i), i = 1, ..., N-1$ is independent from each other. Hence the second term in (133) can be expressed as:

$$H\left(X^{(N-1)n}\right) = \sum_{i=1}^{N-1} H(X^n(i)) \tag{134}$$

From the codebook used by the source node, we observe that

$$H(X^n(i)) = n(B + \min\{C, R_1\}) \tag{135}$$

Therefore

$$H\left(X^{(N-1)n}\right) = (N-1)n(B + \min\{C, R_1\}) \tag{136}$$

The first term in (133) can be upper bounded as:

$$H\left(X^{(N-1)n}|c^{N-1},Y_r^{(N-1)n},X_r^{(N-1)n}\right) \tag{137}$$

$$=\sum_{i=1}^{N-1} H\left(X^n(i)|X^{(i-1)n},c^{N-1},Y_r^{(N-1)n},X_r^{(N-1)n}\right) \tag{138}$$

$$\leq \sum_{i=1}^{N-1} H\left(X^n(i)|c_i,Y_r^n(i),X_r^n(i)\right) \tag{139}$$

Recall that $X^n(i)$ is determined by two indexes: $b_i$ and $c_i$. The label $c_i$ is already on the condition term. To determine $b_i$, we notice from (119) that:

$$B \leq I(X;Y_r|X_r) \tag{140}$$

With this constraint, for a given codebook $\mathcal{C}$, which is on the condition term implicitly, the eavesdropper can estimate $b_i$ from the following set:

$$\{b : X^n(b,c_i), Y_r^n(i), X_r^n(i) \text{ are jointly typical}\} \tag{141}$$

This set should contain only $b_k$ with probability close to 1. From Fano's inequality, we have $H\left(X^n(i)|c_i,Y_r^n(i),X_r^n(i)\right) \leq n\varepsilon_1$, where $\varepsilon_1 > 0$ and $\lim_{n\to\infty}\varepsilon_1 = 0$. Therefore from (137)-(139), we have:

$$H\left(X^{(N-1)n}|c^{N-1},Y_r^{(N-1)n},X_r^{(N-1)n}\right) < (N-1)n\varepsilon_1 \tag{142}$$

The third term in (133) can be upper bounded as follows:

$$I\left(X^{(N-1)n};Y_r^{(N-1)n},X_r^{(N-1)n}\right) \tag{143}$$

$$=\sum_{i=1}^{N-1} I\left(X^{(N-1)n};Y_r^n\left(i\right),X_r^n\left(i\right)|Y_r^{(i-1)n},X_r^{(i-1)n}\right) \tag{144}$$

$$=\sum_{i=1}^{N-1} I\left(X^{(N-1)n};Y_r^n\left(i\right)|Y_r^{(i-1)n},X_r^{(i-1)n},X_r^n\left(i\right)\right)$$

$$+\,I\left(X^{(N-1)n};X_r^n\left(i\right)|Y_r^{(i-1)n},X_r^{(i-1)n}\right) \tag{145}$$

For compress-and-forward relaying, as explained in the previous section, $X_r^n(i)$ is a deterministic function of $Y_r^{(i-1)n},X_r^{(i-1)n}$ . Hence the second term in (145) is zero. (145) therefore equals:

$$\sum_{i=1}^{N-1} I\left(X^{(N-1)n};Y_r^n\left(i\right)|Y_r^{(i-1)n},X_r^{(i-1)n},X_r^n\left(i\right)\right) \tag{146}$$

$$\leq \sum_{i=1}^{N-1} h\left(Y_r^n\left(i\right)|X_r^n\left(i\right)\right) - h\left(Y_r^n\left(i\right)|Y_r^{(i-1)n},X_r^{(i-1)n},X_r^n\left(i\right),X^{(N-1)n}\right) \tag{147}$$

From the coding scheme described in Section A-C, we observe $Y_r^n(i)$ depends on $\{Y_r^{(i-1)n}$, $X_r^{(i-1)n}$, $X^n(1,...,i-1), X^n(i+1,...n)\}$ only through $X_r^n(i), X^n(i)$. Hence

$$Y_r^n\left(i\right)-\left\{X_r^n\left(i\right),X^n\left(i\right)\right\}-\left\{Y_r^{(i-1)n},X_r^{(i-1)n},X^n\left(1,...,i-1\right),X^n\left(i+1,...,n\right)\right\} \tag{148}$$

is a Markov chain. Therefore (147) equals:

$$\sum_{i=1}^{N-1} h\left(Y_r^n\left(i\right)|X_r^n\left(i\right)\right)-h\left(Y_r^n\left(i\right)|X_r^n\left(i\right),X^n\left(i\right)\right) \tag{149}$$

$$=\sum_{i=1}^{N-1} I\left(X^n\left(i\right);Y_r^n\left(i\right)|X_r^n\left(i\right)\right) \tag{150}$$

$$\leq(N-1)(nI(X;Y_r|X_r)+n\varepsilon_2) \tag{151}$$

where where $\varepsilon_2>0$ and $\lim_{n\to\infty}\varepsilon_2=0$. Equation (151) follows from the fact the channel is memoryless and the codebook is composed of i.i.d. sequences.

Applying (142), (136) and (151) to (133), we have

$$H(c^{N-1}|Y_r^{nN},X_r^{nN},\mathcal{C}) \tag{152}$$

$$\geq(N-1)n(B+\min\{C,R_1\})-(N-1)n\varepsilon_1-(N-1)(nI(X;Y_r|X_r)+n\varepsilon_2) \tag{153}$$

$$\geq(N-1)n(\min\{C,R_1\})-(N-1)n(\varepsilon_1+\varepsilon_2) \tag{154}$$

$$=H(c^{N-1}|\mathcal{C})-(N-1)n(\varepsilon_1+\varepsilon_2) \tag{155}$$

Equation (154) follows from (140). (155) is because $c_k$ in each block is chosen independently of other blocks, $c_k$ is chosen according to a uniform distribution from a set with a cardinality of $2^{n\min\{C,R_1\}}$.

From (152)-(155), we have:

$$\lim_{n\to\infty}\frac{1}{n}I(c^{N-1};Y_r^{nN},X_r^{nN}|\mathcal{C})=0 \tag{156}$$

Combining it with (126) we have

$$\lim_{n\to\infty}E[\Pr(\hat{W}^{N-1}\neq W^{N-1}|\mathcal{C})]+\frac{1}{n}I(c^{N-1};Y_r^{nN},X_r^{nN}|\mathcal{C})=0 \tag{157}$$

Therefore, there must exists a codebook $\mathcal{C}^*$ such that

$$\lim_{n\to\infty}E[\Pr(\hat{W}^{N-1}\neq W^{N-1}|\mathcal{C}^*)]+\frac{1}{n}I(c^{N-1};Y_r^{nN},X_r^{nN}|\mathcal{C}^*)=0 \tag{158}$$

Since each term on the left side of (158) are nonnegative, it follows that with this codebook:

$$\lim_{n\to\infty} E[\Pr(\hat{W}^{N-1} \neq W^{N-1}|\mathcal{C}^*)] = 0 \tag{159}$$

$$\lim_{n\to\infty} \frac{1}{n} I(c^{N-1}; Y_r^{nN}, X_r^{nN}|\mathcal{C}^*) = 0 \tag{160}$$

For the simplicity of notation, we omit $\mathcal{C}^*$ from the conditioning term. It is understood that all the derivations below are conditioned on $\mathcal{C}^*$.

Since $c^{N-1}$ is a deterministic function of $W^{N-1}$, we have

$$\begin{aligned}
&H(W^{N-1}|Y_r^{nN}, X_r^{nN}, \mathcal{C}) \\
=&H(c^{N-1}, W^{N-1}|Y_r^{nN}, X_r^{nN}, \mathcal{C}) && (161)\\
\geq&H(c^{N-1}|Y_r^{nN}, X_r^{nN}, \mathcal{C}) && (162)\\
\geq&(N-1)n(\min\{C, R_1\}) - (N-1)n(\varepsilon_1 + \varepsilon_2) && (163)
\end{aligned}$$

Hence,

$$\begin{aligned}
R_e &= \lim_{N\to\infty} \lim_{n\to\infty} \frac{1}{nN} H\left(W^{N-1}|Y_r^{nN}, X_r^{nN}\right) && (164)\\
&\geq \min\left\{\lim_{n\to\infty} C, R_1\right\} && (165)
\end{aligned}$$

The achievable region can then be discussed for two cases:

i) $R_1 \leq \lim_{n\to\infty} C$

ii) $R_1 > \lim_{n\to\infty} C$

When $R_1 \leq \lim_{n\to\infty} C$, the region $(R_1, R_e)$ is given by:

$$\begin{aligned}
&0 \leq R_e \leq R_1 \\
&0 \leq R_1 \leq \lim_{n\to\infty} C
\end{aligned} \tag{166}$$

where $\lim_{n\to\infty} C$ meets the following condition from (125):

$$0 \leq \lim_{n\to\infty} C < [I\left(X; Y, \hat{Y}_r|X_r\right) - I(X; Y_r|X_r)]^+ \tag{167}$$

subject to the constraint $I(X_r; Y) > I(\hat{Y}_r; Y_r|Y X_r)$. When $R_1 > \lim_{n\to\infty} C$, the region $(R_1, R_e)$ is given by:

$$\begin{aligned}
&0 \leq R_e \leq \lim_{n\to\infty} C \\
&\lim_{n\to\infty} C < R_1 \leq \lim_{n\to\infty} C + I(X; Y_r|X_r)
\end{aligned} \tag{168}$$

Finally, the union of these two regions (166) and (168) becomes the region given below:

$$\begin{aligned} R_e \leq R_1 &< I\left(X;Y,\hat{Y}_r|X_r\right) \\ 0 \leq R_e &< [I\left(X;Y,\hat{Y}_r|X_r\right) - I\left(X;Y_r|X_r\right)]^+ \end{aligned} \tag{169}$$

with the constraint $I(X_r;Y) > I(\hat{Y}_r;Y_r|YX_r)$.

## Appendix B
## Proof of Theorem 2

The converse for $R_1$ is given in [22] using the cut set bound. The converse for $R_e$ can be derived by specializing the upper bound in [21], which is stated in (58) as $I(X;Y|X_r,Y_r)$. For our model, this can be upper bounded as:

$$I\left(X;Y|Y_r,X_r\right) = I\left(X_R,X_D;Y|Y_r,X_r\right) \tag{170}$$

$$= h\left(Y|Y_r,X_r\right) - h\left(Y|Y_r,X_r,X_R,X_D\right) \tag{171}$$

From (12), $Y - X_D, X_r - Y_r, X_R$ is a Markov chain. Hence (171) equals:

$$h\left(Y|Y_r,X_r\right) - h\left(Y|X_r,X_D\right) \tag{172}$$

$$\leq h\left(Y|X_r\right) - h\left(Y|X_r,X_D\right) \tag{173}$$

$$= I\left(X_D;Y|X_r\right) \tag{174}$$

Hence we have proved the converse for $R_e$.

The achievability of (21) also follows from the partial decode-and-forward scheme presented in reference [21, Theorem 1].

*Theorem 8:* [21, Theorem 1] The following region is achievable:

$$\bigcup_{\substack{p(U,X,X_r)\\ p(Y_r,Y|X,X_r)}} \left\{ \begin{array}{l} R_1 < I\left(X;Y|U,X_r\right) + \min\{I\left(U;Y_r|X_r\right), I\left(U,X_r;Y\right)\} \\ R_e < [I\left(X;Y|U,X_r\right) - I\left(X;Y_r|U,X_r\right)]^+ \\ 0 \leq R_e \leq R_1 \end{array} \right\} \tag{175}$$

In (175), we let $X = \{X_D, X_R\}, U = X_R$, and restrict the union to be over the probability distributions of the form $p(X_r)p(X_D|X_r)p(X_R|X_r)$, and we obtain:

$$I\left(X;Y|U,X_r\right) - I\left(X;Y_r|U,X_r\right) \tag{176}$$

$$= I\left(X_RX_D;Y|X_RX_r\right) - I\left(X_RX_D;Y_r|X_RX_r\right) \tag{177}$$

$$= I\left(X_D;Y|X_RX_r\right) - I\left(X_D;Y_r|X_RX_r\right) \tag{178}$$

$$=H\left(Y|X_RX_r\right)-H\left(Y|X_DX_RX_r\right)-I\left(X_D;Y_r|X_RX_r\right) \tag{179}$$

$$\overset{(a)}{=}H\left(Y|X_r\right)-H\left(Y|X_DX_r\right)-I\left(X_D;Y_r|X_RX_r\right) \tag{180}$$

$$=I\left(X_D;Y|X_r\right)-I\left(X_D;Y_r|X_RX_r\right) \tag{181}$$

$$\overset{(b)}{=}I\left(X_D;Y|X_r\right) \tag{182}$$

where step $(a)$ follows from $X_R-X_r-Y$ being a Markov chain [22] and $X_R-X_rX_D-Y$ being a Markov chain. Step $(b)$ follows from $X_D-X_RX_r-Y_r$ being a Markov chain [22]. Moreover, the bound on $R_1$ can be expressed as:

$$I\left(X;Y|U,X_r\right)+\min\{I\left(U;Y_r|X_r\right),I\left(U,X_r;Y\right)\} \tag{183}$$

$$=\min\left\{\ I\left(UXX_r;Y\right),I\left(U;Y_r|X_r\right)+I\left(X;Y|U,X_r\right)\ \right\} \tag{184}$$

$$\overset{(c)}{=}\min\left\{\ I\left(XX_r;Y\right),I\left(U;Y_r|X_r\right)+I\left(X;Y|U,X_r\right)\ \right\} \tag{185}$$

where step $(c)$ follows from $U-XX_r-Y$ being a Markov chain.

Note that (185) is the same as [22, (2)], therefore from the same argument therein, we obtain:

$$\min\left\{\ I\left(XX_r;Y\right),I\left(U;Y_r|X_r\right)+I\left(X;Y|U,X_r\right)\ \right\} \tag{186}$$

$$=\min\left\{\ I\left(X_D,X_r;Y\right),I\left(X_R;Y_r|X_r\right)+I\left(X_D;Y|X_r\right)\ \right\} \tag{187}$$

By substituting (187) and (182) into (175), we find that the rate pair in (21) is achievable.

*Remark 17:* It is shown in [21, Lemma 3] that the achievable rate region (175) is convex. Therefore the rate region (21) is also convex. □

## Appendix C

### Proof that (38) is a Monotonic Increasing Function of the Source Power

It suffices to show that the argument of $C(\ \ )$, which is $p+\frac{a^2p}{1+\sigma_Q^2}$, is a monotonically increasing function of $p$. The expression of $\sigma_Q^2$ is given by (32). Thus we have:

$$p+\frac{a^2p}{1+\sigma_Q^2}=p\left(\frac{(1+a^2)\,(b^2P_r+1)\,(p+1)-a^2}{(b^2P_r+a^2+1)\,(p+1)-a^2}\right) \tag{188}$$

Since $p+\frac{a^2p}{1+\sigma_Q^2}$ is always positive, we can prove its monotonicity in $p$ by showing $\ln(p+\frac{a^2p}{1+\sigma_Q^2})$ is monotonically increasing in $p$. From (188), it is given by

$$\ln p+\ln\left(\left(1+a^2\right)\left(b^2P_r+1\right)(p+1)-a^2\right)-\ln\left(\left(b^2P_r+a^2+1\right)(p+1)-a^2\right) \tag{189}$$

Let $A \triangleq (1+a^2)(b^2P_r+1)$. Let $B \triangleq (b^2P_r+a^2+1)$. Then the derivative of (189) with respect to $p$ is given by

$$\frac{1}{p}+\frac{a^2\left(\frac{1}{A}-\frac{1}{B}\right)}{\left(p+1-\frac{a^2}{A}\right)\left(p+1-\frac{a^2}{B}\right)}=\frac{\left(p+1-\frac{a^2}{A}\right)\left(p+1-\frac{a^2}{B}\right)+a^2\left(\frac{1}{A}-\frac{1}{B}\right)p}{p\left(p+1-\frac{a^2}{A}\right)\left(p+1-\frac{a^2}{B}\right)} \tag{190}$$

Note that $A, B > a^2$. Hence the denominator of (190) is positive. Therefore we only need to show that the numerator of (190) is positive. The numerator of (190) equals

$$(p+1)^2-a^2\left(\frac{1}{A}+\frac{1}{B}\right)(p+1)+\frac{a^4}{AB}+a^2\left(\frac{1}{A}-\frac{1}{B}\right)p \tag{191}$$

$$=(p+1)^2-\frac{2a^2}{B}p+\frac{a^4}{AB}-a^2\left(\frac{1}{A}+\frac{1}{B}\right) \tag{192}$$

$$=\left(p+1-\frac{a^2}{B}\right)^2-\frac{a^4}{B}\left(\frac{1}{B}-\frac{1}{A}\right)+a^2\left(\frac{1}{B}-\frac{1}{A}\right) \tag{193}$$

$$=\left(p+1-\frac{a^2}{B}\right)^2+a^2\left(1-\frac{a^2}{B}\right)\left(\frac{1}{B}-\frac{1}{A}\right) \tag{194}$$

Since $A > B > a^2$, $a^2/B < 1$, (194) is positive. Therefore the derivative of $\ln(p+\frac{a^2p}{1+\sigma_Q^2})$ with respect to $p$ is positive. This means (38) is a monotonic increasing function of $p$.

## Appendix D

## An Example where introducing a Second Eavesdropper Decreases Secrecy Capacity

Consider a special case of Model $2$ defined by

$$Y_r = X+N \quad Y_D = X-N \tag{195}$$

$$Y_R = X_r \tag{196}$$

This is a Gaussian relay channel with orthogonal components with reversely correlated noise. $N$ is a zero mean Gaussian random variable with unit variance. Hence its probability density function is symmetric around the origin: $p(-N)=p(N)$.

We first observe that since the orthogonal link between the relay and the destination is noiseless, the optimal relaying scheme in this case is choosing $X_{r,i}=Y_{r,i-1}$. This can be proved as follows: First we recognize, for this channel, given $Y_r^n$, the signals $X_r^n$ do not provide more information to the eavesdropper. This is because the relay is not interfering itself and hence as shown in Remark 11, the secrecy capacity can be computed from $\lim_{n\to\infty}\frac{1}{n}H(W|Y_r^n)$ instead, i.e., $X_r^n$ can be dropped from the conditioning term. Therefore for any given relay scheme, we can always use $X_{r,i} = Y_{r,i-1}$ to give the destination the signals received by the relay, and ask the destination to to compute the $X_r^n$ generated from the original relaying scheme instead. It can be verified that in this way the secrecy constraint is fulfilled and $W$ can still be transmitted reliably. Therefore, the secrecy rate achievable by any given relay scheme is achievable via $X_{r,i} = Y_{r,i-1}$, which must be the optimal relaying scheme.

Hence the destination essentially receives $(Y_{r,i-1}, Y_{D,i})$ at the $i$th channel use and the eavesdropper receives $Y_r$. The channel is therefore equivalent to a $1 \times 2$ MIMO wiretap channel [4]. Note that the destination can remove the noise $N_i$ completely by simply computing $Y_{r,i} + Y_{D,i}$. The eavesdropper, on the other hand, observes an AWGN link with finite capacity. Hence the secrecy capacity of this channel is easily seen to be $\infty$.

Now, we construct a second relay channel. The channel is the same as the previous one except that the received signal at the relay becomes:

$$\{Y_r, X_r\} = \{X + N, X_r\} \tag{197}$$

That is to say that the relay receives an additional copy of its transmitted signal. This should not benefit the relay/eavesdropper at all. So the secrecy capacity is still $\infty$ .

Now, we construct a third relay channel from the second relay channel, by adding one more eavesdropper to the model. Let the signal received by this second eavesdropper be:

$$Y_e = \{X - N, X_r\} \tag{198}$$

It follows that $p(Y_e|X, X_r) = p(X - N, X_r|X, X_r) = p(-N) = p(N) = p(Y_r|X, X_r)$. Hence, the new eavesdropper observes the same marginal distribution as the eavesdropper located at the relay node. However, this eavesdropper receives exactly the same signal received by the destination. Therefore the secrecy capacity of the new system is reduced to 0.

## Appendix E
## Proof that (45) as a Markov Chain

Recall that $A$ is shown in (1) as the local randomness generated at the relay. Using chain rule, we have

$$\begin{aligned}&p\left(Y_r^n, Y_e^n, X^n, X_r^n, A\right)\\&= p\left(X^n, A\right)\prod_{i=1}^{n} p\left(X_{r,i}|Y_{r,1}^{i-1}, X_{r,1}^{i-1}, A, Y_{e,1}^{i-1}, X^n\right) p\left(Y_{e,i}, Y_{r,i}|Y_{e,1}^{i-1}, Y_{r,1}^{i-1}, X_{r,1}^{i}, X^n, A\right)\end{aligned} \tag{199}$$

From (1), we observe that

$$X_{r,i} - \{Y_{r,1}^{i-1}, X_{r,1}^{i-1}, A\} - \{Y_{e,1}^{i-1}, X^n\} \tag{200}$$

is a Markov chain. Since the channel is memoryless, and the relay function (1) has to be causal, we observe

$$\{Y_{e,i}, Y_{r,i}\} - \{X_{r,i}, X_i\} - \{Y_{e,1}^{i-1}, Y_{r,1}^{i-1}, X_{r,1}^{i-1}, X_{]i[}, A\} \tag{201}$$

is also a Markov chain. Applying these two Markov chains to (199), we have:

$$p\left(Y_r^n, Y_e^n, X^n, X_r^n, A\right) = p\left(X^n, A\right)\prod_{i=1}^{n} p\left(X_{r,i}|Y_{r,1}^{i-1}, X_{r,1}^{i-1}, A\right) p\left(Y_{e,i}, Y_{r,i}|X_{r,i}, X_i\right) \tag{202}$$

We next integrate out $Y_r^n$ and $X_r^n$ from both side of (202). This can be done in a recursive fashion as we show next. First we integrate over $Y_{r,n}$ on both side of (202). This gives us:

$$\begin{aligned}&p\left(Y_r^{n-1}, Y_e^n, X^n, X_r^n, A\right)\\=&p\left(X^n, A\right)\prod_{i=1}^{n-1} p\left(X_{r,i}|Y_{r,1}^{i-1}, X_{r,1}^{i-1}, A\right) p\left(Y_{e,i}, Y_{r,i}|X_{r,i}, X_i\right)\\&p\left(X_{r,n}|Y_{r,1}^{n-1}, X_{r,1}^{n-1}, A\right) p\left(Y_{e,n}|X_{r,n}, X_n\right)\end{aligned} \tag{203}$$

From (43), we have $p(Y_{e,n}|X_{r,n}, X_n) = p(Y_{e,n}|X_n)$. Applying it to (203) yields:

$$\begin{aligned}&p\left(Y_r^{n-1}, Y_e^n, X^n, X_r^n, A\right)\\=&p\left(X^n, A\right)\prod_{i=1}^{n-1} p\left(X_{r,i}|Y_{r,1}^{i-1}, X_{r,1}^{i-1}, A\right) p\left(Y_{e,i}, Y_{r,i}|X_{r,i}, X_i\right)\\&p\left(X_{r,n}|Y_{r,1}^{n-1}, X_{r,1}^{n-1}, A\right) p\left(Y_{e,n}|X_n\right)\end{aligned} \tag{204}$$

We next integrate over $X_{r,n}$ on both side of (204), which yields:

$$\begin{aligned} & p\left(Y_{r,1}^{n-1}, Y_e^n, X^n, X_{r,1}^{n-1}, A\right) \\ =& p\left(X^n, A\right) p\left(Y_{e,n}|X_n\right) \prod_{i=1}^{n-1} p\left(X_{r,i}|Y_{r,1}^{i-1}, X_{r,1}^{i-1}, A\right) p\left(Y_{e,i}, Y_{r,i}|X_{r,i}, X_i\right) \end{aligned} \tag{205}$$

Repeating the process in (203)-(205) for $n-1, n-2, ..., 1$, we have

$$p\left(Y_e^n, X^n, A\right) = p\left(X^n, A\right) \prod_{i=1}^{n} p\left(Y_{e,i}|X_i\right) \tag{206}$$

Integrating over $A$ on both sides of (206), we have

$$p\left(Y_e^n, X^n\right) = p\left(X^n\right) \prod_{i=1}^{n} p\left(Y_{e,i}|X_i\right) \tag{207}$$

Integrating over $Y_{e,j+1}^n$ on both sides of (207), we have

$$p\left(Y_{e,1}^j, X^n\right) = p\left(X^n\right) \prod_{i=1}^{j} p\left(Y_{e,i}|X_i\right) \tag{208}$$

Integrating over $Y_{e,j}$ on both sides of (208), we have

$$p\left(Y_{e,1}^{j-1}, X^n\right) = p\left(X^n\right) \prod_{i=1}^{j-1} p\left(Y_{e,i}|X_i\right) \tag{209}$$

Dividing each side of (208) by the corresponding side of (209), we have

$$p\left(Y_{e,j}|Y_{e,1}^{j-1}, X^n\right) = p\left(Y_{e,j}|X_j\right) \tag{210}$$

Hence we have shown that $Y_{e,j} - X_j - X_{]j[}Y_{e,1}^{j-1}$ is a Markov chain.

We next prove that $Y_{r,j} - X_j - X_{]j[}Y_{r,1}^{j-1}$ is a Markov chain. Again, we start with (202) and integrate out $Y_e^n$ and $X_r^n$ from both side of it in a recursive fashion. First we integrate over $Y_{e,n}$ on both side of (202) and obtain

$$\begin{aligned} & p\left(Y_r^n, Y_{e,1}^{n-1}, X^n, X_r^n, A\right) \\ =& p\left(X^n, A\right) \prod_{i=1}^{n-1} p\left(X_{r,i}|Y_{r,1}^{i-1}, X_{r,1}^{i-1}, A\right) p\left(Y_{e,i}, Y_{r,i}|X_{r,i}, X_i\right) \\ & p\left(X_{r,n}|Y_{r,1}^{n-1}, X_{r,1}^{n-1}, A\right) p\left(Y_{r,n}|X_{r,n}, X_n\right) \end{aligned} \tag{211}$$

Then from (43) we observe that $p\left(Y_{r,n}|X_{r,n}, X_n\right) = p\left(Y_{r,n}|X_n\right)$. Hence (211) becomes:

$$\begin{aligned} & p\left(Y_r^n, Y_{e,1}^{n-1}, X^n, X_r^n, A\right) \\ =& p\left(X^n, A\right) \prod_{i=1}^{n-1} p\left(X_{r,i}|Y_{r,1}^{i-1}, X_{r,1}^{i-1}, A\right) p\left(Y_{e,i}, Y_{r,i}|X_{r,i}, X_i\right) \\ & p\left(X_{r,n}|Y_{r,1}^{n-1}, X_{r,1}^{n-1}, A\right) p\left(Y_{r,n}|X_n\right) \end{aligned} \tag{212}$$

We next integrate over $X_{r,n}$ on both side of (212), which yields:

$$\begin{aligned}&p\left(Y_r^n, Y_{e,1}^{n-1}, X^n, X_r^{n-1}, A\right)\\&=p\left(X^n, A\right) p\left(Y_{r,n}|X_n\right) \prod_{i=1}^{n-1} p\left(X_{r,i}|Y_{r,1}^{i-1}, X_{r,1}^{i-1}, A\right) p\left(Y_{e,i}, Y_{r,i}|X_{r,i}, X_i\right)\end{aligned} \tag{213}$$

Repeating the process in (211)-(213) for $n-1, n-2, ..., 1$, we have

$$p\left(Y_r^n, X^n, A\right) = p\left(X^n, A\right) \prod_{i=1}^{n} p\left(Y_{r,i}|X_i\right) \tag{214}$$

Integrating over $A$ on both sides of (214), we have

$$p\left(Y_r^n, X^n\right) = p\left(X^n\right) \prod_{i=1}^{n} p\left(Y_{r,i}|X_i\right) \tag{215}$$

Integrating over $Y_{r,j+1}^n$ on both sides of (215), we have

$$p\left(Y_{r,1}^j, X^n\right) = p\left(X^n\right) \prod_{i=1}^{j} p\left(Y_{r,i}|X_i\right) \tag{216}$$

Integrating over $Y_{r,j}$ on both sides of (216), we have

$$p\left(Y_{r,1}^{j-1}, X^n\right) = p\left(X^n\right) \prod_{i=1}^{j-1} p\left(Y_{r,i}|X_i\right) \tag{217}$$

Dividing each side of (216) by the corresponding side of (217), we have

$$p\left(Y_{r,j}|Y_{r,1}^{j-1}, X^n\right) = p\left(Y_{r,j}|X_j\right) \tag{218}$$

Hence we have shown that $Y_{r,j} - X_j - X_{]j[}Y_{r,1}^{j-1}$ is a Markov chain.

## Appendix F

## Proof that $H(W|Y_r^n X_r^n) = H(W|Y_r^n)$ for the Model stated in (41)

We begin with

$$H(W|Y_r^n, X_r^n) = H(W|Y_r^n, X_r^n, A) = H(W|Y_r^n, A) \tag{219}$$

where the last equality follows from the fact that $X_r^n$ is a deterministic function of $\{Y_r^n, A\}$. Hence we only need to prove $H(W|Y_r^n, A) = H(W|Y_r^n)$ for the channel model defined in (41). This can be done as follows:

First we factorize $p\left(Y_r^n, X^n, X_r^n, A, W\right)$ using a similar procedure seen in (199)-(202):

$$p\left(Y_r^n, X^n, X_r^n, A, W\right) \tag{220}$$

$$=p\left(W, A, X^n\right) p\left(Y_r^n, X_r^n | X^n, A, W\right) \tag{221}$$

$$=p\left(W, A, X^n\right) \prod_{i=1}^{n} p\left(X_{r,i} | Y_{r,1}^{i-1}, X_{r,1}^{i-1}, A\right) p\left(Y_{r,i} | X_{r,i}, X_i\right) \tag{222}$$

where in (222) we use the Markov chain stated in (200) and (201).

From (41), we have $p\left(Y_{r,i} | X_{r,i}, X_i\right) = p\left(Y_{r,i} | X_i\right)$. Hence we have

$$\begin{aligned} & p\left(Y_r^n, X^n, X_r^n, A, W\right) \\ =& p\left(W, A, X^n\right) \prod_{i=1}^{n} p\left(X_{r,i} | Y_{r,1}^{i-1}, X_{r,1}^{i-1}, A\right) p\left(Y_{r,i} | X_i\right) \end{aligned} \tag{223}$$

We next integrate out $X_r^n$ from both sides of (223) using the procedure shown in Appendix E, which yields:

$$p\left(Y_r^n, X^n, A, W\right) = p\left(W, X^n, A\right) \prod_{i=1}^{n} p\left(Y_{r,i} | X_i\right) \tag{224}$$

We next use the fact that $Y_{r,i} - X_i - X_{]i[} Y_{r,1}^{i-1}$ is a Markov chain, as stated in (45) and proved in Appendix E, from which we have

$$p\left(Y_r^n, X^n, A, W\right) = p\left(W, X^n, A\right) \prod_{i=1}^{n} p\left(Y_{r,i} | X^n, Y_{r,1}^{i-1}\right) \tag{225}$$

Since $W$ is a deterministic function of $X^n$, we have

$$p\left(Y_r^n, X^n, A, W\right) = p\left(W, X^n, A\right) \prod_{i=1}^{n} p\left(Y_{r,i} | X^n, Y_{r,1}^{i-1}, W\right) \tag{226}$$

$$= p\left(W, X^n, A\right) p\left(Y_r^n | X^n, W\right) \tag{227}$$

Since $A$ is independent from $W, X^n$, (226)-(227) can be written as

$$p\left(Y_r^n, X^n, A, W\right) = p\left(W, X^n, Y_r^n\right) p\left(A\right) \tag{228}$$

From it, we can write:

$$p\left(Y_r^n, A\right) = p\left(Y_r^n\right) p\left(A\right) \tag{229}$$

and

$$p\left(Y_r^n, A, W\right) = p\left(W, Y_r^n\right) p\left(A\right) \tag{230}$$

From (229) and (230), we have

$$p\left(W|Y_r^n, A\right) = p\left(W|Y_r^n\right) \tag{231}$$

Hence $H(W|Y_r^n, A) = H(W|Y_r^n)$.

## Appendix G
## Proof that (79) is an Upper Bound

We begin with

$$H\left(W\right) = H\left(W|Y_e^n\right) + I\left(W;Y_e^n\right) \tag{232}$$

Due to the secrecy constraint, we have $\lim_{n\to\infty}\frac{1}{n}I(W;Y_e^n) = 0$. Due to the fact that $W$ can be decoded from $Y_D^n, Y_R^n$ reliably, we have, from Fano's inequality, $\lim_{n\to\infty}\frac{1}{n}H\left(W|Y_D^nY_R^n\right) = 0$. Hence there exists $\varepsilon > 0$ such that

$$I\left(W;Y_e^n\right) < n\varepsilon/2 \tag{233}$$

$$H\left(W|Y_D^nY_R^n\right) < n\varepsilon/2 \tag{234}$$

$$\lim_{n\to\infty}\varepsilon = 0 \tag{235}$$

For this $\varepsilon$, we find (232) is upper bounded by

$$H\left(W|Y_e^n\right) + n\varepsilon/2 \tag{236}$$

$$\leq H\left(W|Y_e^n\right) - H\left(W|Y_D^nY_R^n\right) + n\varepsilon \tag{237}$$

$$\leq H\left(W|Y_e^n\right) - H\left(W|Y_D^nY_R^nX_r^n\right) + n\varepsilon \tag{238}$$

$$\leq I\left(W;Y_D^n, Y_R^n, X_r^n|Y_e^n\right) + n\varepsilon \tag{239}$$

$$\leq I\left(W, X^n;Y_D^n, Y_R^n, X_r^n|Y_e^n\right) + n\varepsilon \tag{240}$$

$$= I\left(X^n;Y_D^n, Y_R^n, X_r^n|Y_e^n\right) + n\varepsilon \tag{241}$$

$$= I\left(X^n;Y_D^n, X_r^n|Y_e^n\right) + I\left(X^n;Y_R^n|Y_e^n, Y_D^n, X_r^n\right) + n\varepsilon \tag{242}$$

$$= I\left(X^n;Y_D^n, X_r^n|Y_e^n\right) + \sum_{i=1}^{n} I\left(X^n;Y_{R,i}|Y_e^n, Y_D^n, X_r^n, Y_{R,1}^{i-1}\right) + n\varepsilon \tag{243}$$

From (55), we observe $Y_{R,i} - \{Y_e^n, Y_D^n, X_r^n, Y_{R,1}^{i-1}\} - X^n$ is a Markov chain. Hence (243) equals:

$$I\left(X^n; Y_D^n, X_r^n | Y_e^n\right) + n\varepsilon \tag{244}$$

$$\leq I\left(X^n; Y_D^n, Y_r^n, X_r^n | Y_e^n\right) + n\varepsilon \tag{245}$$

$$= I\left(X^n; Y_D^n, Y_r^n | Y_e^n\right) + I\left(X^n; X_r^n | Y_D^n, Y_r^n, Y_e^n\right) + n\varepsilon \tag{246}$$

$$= I\left(X^n; Y_D^n, Y_r^n | Y_e^n\right) + \sum_{i=1}^{n} I\left(X^n; X_{r,i} | X_{r,1}^{i-1}, Y_D^n, Y_r^n, Y_e^n\right) + n\varepsilon \tag{247}$$

The term inside the sum in (247) can be bounded as

$$I\left(X^n; X_{r,i} | X_{r,1}^{i-1}, Y_D^n, Y_r^n, Y_e^n\right) \tag{248}$$

$$\leq I\left(X^n; A, X_{r,1}^{i} | Y_D^n, Y_r^n, Y_e^n\right) \tag{249}$$

$$\leq I\left(X^n; A | Y_D^n, Y_r^n, Y_e^n\right) + I\left(X^n; X_{r,1}^{i} | Y_D^n, Y_r^n, Y_e^n, A\right) \tag{250}$$

$$= I\left(X^n; A | Y_D^n, Y_r^n, Y_e^n\right) \tag{251}$$

where $A$ is the local randomness at the relay. (251) is due to the fact that $X_{r,1}^{i}$ is a deterministic function of $\{Y_r^n, A\}$.

From (55), we have

$$p\left(X^n, X_r^n, A, Y_D^n, Y_r^n, Y_e^n\right) = p\left(X^n\right) p\left(A\right) \prod_{i=1}^{n} p\left(Y_{D,i}, Y_{r,i}, Y_{e,i} | X_i\right) p\left(X_{r,i} | X_{r,1}^{i-1}, Y_{r,1}^{i-1}, A\right) \tag{252}$$

from which we have

$$p\left(X^n, Y_D^n, Y_r^n, Y_e^n, A\right) = p\left(X^n\right) p\left(A\right) \prod_{i=1}^{n} p\left(Y_{D,i}, Y_{r,i}, Y_{e,i} | X_i\right) \tag{253}$$

Hence $\{Y_D^n, Y_r^n, Y_e^n, X^n\}$ are all independent from $A$. Therefore (251) equals $0$. (247) thus becomes:

$$I\left(X^n; Y_D^n, Y_r^n | Y_e^n\right) + n\varepsilon \tag{254}$$

$$= \sum_{i=1}^{n} h\left(Y_{D,i} Y_{r,i} | Y_{D,1}^{i-1} Y_{r,1}^{i-1} Y_e^n\right) - \sum_{i=1}^{n} h\left(Y_{D,i} Y_{r,i} | Y_{D,1}^{i-1} Y_{r,1}^{i-1} Y_e^n X^n\right) + n\varepsilon \tag{255}$$

$$\leq \sum_{i=1}^{n} h\left(Y_{D,i} Y_{r,i} | Y_{e,i}\right) - \sum_{i=1}^{n} h\left(Y_{D,i} Y_{r,i} | Y_{D,1}^{i-1} Y_{r,1}^{i-1} Y_e^n X^n\right) + n\varepsilon \tag{256}$$

From (55) and the fact that the channel is memoryless and the relay function is causal, we observe that

$$\{Y_{D,i}, Y_{r,i}\} - \{Y_{e,i}, X_i\} - \{Y_{D,1}^{i-1}, Y_{r,1}^{i-1}, Y_{e,]i[}, X_{]i[}\} \tag{257}$$

is a Markov chain. Hence (256) equals:

$$\sum_{i=1}^{n} h\left(Y_{D,i}Y_{r,i}|Y_{e,i}\right) - \sum_{i=1}^{n} h\left(Y_{D,i}Y_{r,i}|Y_{e,i}X_i\right) + n\varepsilon \tag{258}$$

$$= \sum_{i=1}^{n} I\left(X_i; Y_{D,i}Y_{r,i}|Y_{e,i}\right) + n\varepsilon \tag{259}$$

Define $Q$ as a random variable that is uniformly distributed over $\{1, 2, ..., n\}$. Define $X = X_Q, Y_D = Y_{D,Q}, Y_r = Y_{r,Q}, Y_e = Y_{e,Q}$. Then (259) equals:

$$nI\left(X; Y_D Y_r|Y_e Q\right) + n\varepsilon \tag{260}$$

$$= n\left(h\left(Y_D Y_r|Y_e Q\right) - h\left(Y_D Y_r|Y_e X Q\right)\right) + n\varepsilon \tag{261}$$

$$\leq n\left(h\left(Y_D Y_r|Y_e\right) - h\left(Y_D Y_r|Y_e X Q\right)\right) + n\varepsilon \tag{262}$$

Since $\{Y_D, Y_r\} - \{Y_e, X\} - Q$ is a Markov chain, (262) equals:

$$n\left(h\left(Y_D Y_r|Y_e\right) - h\left(Y_D Y_r|Y_e X\right)\right) + n\varepsilon \tag{263}$$

$$= nI\left(X; Y_D Y_r|Y_e\right) + n\varepsilon \tag{264}$$

Dividing both sides by $n$ and letting $n \to \infty$, we have the upper bound in (79).

## References

[1] C. E. Shannon. Communication Theory of Secrecy Systems. *Bell System Technical Journal*, 28(4):656–715, September 1949.

[2] A. D. Wyner. The Wire-tap Channel. *Bell System Technical Journal*, 54(8):1355–1387, 1975.

[3] I. Csiszár and J. Körner. Broadcast Channels with Confidential Messages. *IEEE Transactions on Information Theory*, 24(3):339–348, May 1978.

[4] S. Shafiee, N. Liu, and S. Ulukus. Towards the Secrecy Capacity of the Gaussian MIMO Wire-tap Channel: The 2-2-1 Channel. Submitted to IEEE Transactions on Information Theory, September, 2007.

[5] A. Khisti and G. Wornell. Secure Transmission with Multiple Antennas: The MISOME Wiretap Channel. Submitted to IEEE Transactions on Information Theory, August, 2007.

[6] F. Oggier and B. Hassibi. The Secrecy Capacity of the MIMO Wiretap Channel. In *IEEE International Symposium on Information Theory*, July 2008.

[7] X. He and A. Yener. Providing Secrecy With Structured Codes: Tools and Applications to Gaussian Two-user Channels. Submitted to IEEE Transactions on Information Theory, July, 2009, Available online at http://arxiv.org/abs/0907.5388.

[8] E. Tekin, S. Serbetli, and A. Yener. On Secure Signaling for the Gaussian Multiple Access Wire-tap Channel. In *39th Annual Asilomar Conference on Signals, Systems, and Computers*, November 2005.

[9] E. Tekin and A. Yener. The Gaussian Multiple Access Wire-tap Channel. *IEEE Transactions on Information Theory*, 54(12):5747–5755, December 2008.

[10] E. Tekin and A. Yener. The General Gaussian Multiple Access and Two-Way Wire-Tap Channels: Achievable Rates and Cooperative Jamming. *IEEE Transactions on Information Theory*, 54(6):2735–2751, June 2008.

[11] E. Ekrem and S. Ulukus. On the Secrecy of Multiple Access Wiretap Channel. In *46th Allerton Conference on Communication, Control, and Computing*, September 2008.

[12] E. Ekrem and S. Ulukus. The Secrecy Capacity Region of the Gaussian MIMO Multi-receiver Wiretap Channel. Submitted to IEEE Transactions on Information Theory, March 2009.

[13] L. Lai and H. El Gamal. Cooperation for Secrecy: The Relay-Eavesdropper Channel. *IEEE Transactions on Information Theory*, 54(9):4005–4019, September 2008.

[14] E. Tekin and A. Yener. Achievable Rates for the General Gaussian Multiple Access Wire-Tap Channel with Collective Secrecy. In *Allerton Conference on Communication, Control, and Computing*, September 2006.

[15] Y. Liang and H. V. Poor. Multiple-Access Channels with Confidential Messages. *IEEE Transactions on Information Theory*, 54(3):976–1002, March 2008.

[16] R. Liu and H. V. Poor. Multi-Antenna Gaussian Broadcast Channels with Confidential Messages. In *IEEE International Symposium on Information Theory*, July 2008.

[17] R. Liu, T. Liu, H. V. Poor, and S. Shamai. Multiple-Input Multiple-Output Gaussian Broadcast Channels with Confidential Messages. Submitted to IEEE Transactions on Information Theory, March 2009.

[18] R. Liu, I. Maric, P. Spasojevic, and R. D. Yates. Discrete Memoryless Interference and Broadcast Channels with Confidential Messages: Secrecy Rate Regions. *IEEE Transactions on Information Theory*, 54(6):2493–2507, June 2008.

[19] R. D. Yates, D. Tse, and Z. Li. Secure Communication on Interference Channels. In *IEEE International Symposium on Information Theory*, July 2008.

[20] Y. Oohama. Coding for Relay Channels with Confidential Messages. In *Information Theory Workshop*, September 2001.

[21] Y. Oohama. Relay Channels with Confidential Messages. Submitted to IEEE Transactions on Information Theory, March, 2007.

[22] A. E. Gamal and S. Zahedi. Capacity of a Class of Relay Channels with Orthogonal Components. *IEEE Transactions on Information Theory*, 51(5):1815–1817, May 2005.

[23] Y. Liang and V. V. Veeravalli. Gaussian Orthogonal Relay Channels: Optimal Resource Allocation and Capacity. *IEEE Transactions on Information Theory*, 51(9):3284–3289, September 2005.

[24] Y. H. Kim. Capacity of a Class of Deterministic Relay Channels. *IEEE Transactions on Information Theory*, 54(3):1328–1329, March 2008.

[25] T. Cover and A. E. Gamal. Capacity Theorems for the Relay Channel. *IEEE Transactions on Information Theory*, 25(5):572–584, September 1979.

[26] M. Yuksel and E. Erkip. Secure Communication with a Relay Helping the Wiretapper. In *IEEE Information Theory Workshop*, September 2007.

[27] S. Leung-Yan-Cheong and M. Hellman. The Gaussian Wire-tap Channel. *IEEE Transactions on Information Theory*, 24(4):451–456, July 1978.

[28] T. M. Cover and Y. H. Kim. Capacity of a Class of Deterministic Relay Channels. In *IEEE International Symposium on Information Theory*, June 2008.

[29] X. He and A. Yener. On the Equivocation Region of Relay Channels with Orthogonal Components. In *Annual Asilomar Conference on Signals, Systems, and Computers*, November 2007.

[30] E. Ekrem and S. Ulukus. Effects of Cooperation on the Secrecy of Multiple Access Channels with Generalized Feedback. In *Annual Conference on Information Sciences and Systems*, March 2008.

[31] E. Ekrem and S. Ulukus. Secrecy in Cooperative Relay Broadcast Channels. In *IEEE International Symposium on Information Theory*, July 2008.

[32] E. Ekrem and S. Ulukus. Secrecy in Cooperative Relay Broadcast Channels. Submitted to IEEE Transactions on Information Theory, October, 2008.

[33] X. He and A. Yener. Two-hop Secure Communication Using an Untrusted Relay: A Case for Cooperative Jamming. In *IEEE Global Telecommunication Conference*, November 2008.

[34] X. He and A. Yener. On the Role of Feedback in Two Way Secure Communication. In 42*nd Annual Asilomar Conference on Signals, Systems, and Computers*, October 2008.

[35] X. He and A. Yener. Providing Secrecy with Lattice Codes. In *46th Allerton Conference on Communication, Control, and Computing*, September 2008.

[36] X. He and A. Yener. End-to-end Secure Multihop Communication Using Untrusted Relays is Possible. In 42*nd Annual Asilomar Conference on Signals, Systems, and Computers*, October 2008.